\documentclass[preprint,12pt]{elsarticle}

\usepackage{tikz}
\usepackage{amsmath,bm,algorithm}
\usepackage[mathscr]{euscript}
 \let\mathscr\relax
\usepackage{amssymb}
\usepackage{mathrsfs}
\usepackage{graphicx}

\usepackage{natbib}
\usepackage{lineno,hyperref}
\usepackage{xcolor}
\usepackage{caption}
\usepackage{booktabs}
\usepackage{makecell}

\usepackage{xcolor} 

\usepackage{placeins}  

\usepackage{siunitx}  

\usepackage{subcaption}

\usepackage[noend]{algpseudocode}

\begin{document}

\begin{frontmatter}

\title{Deep learning driven framework for optimization of polycrystalline microstructures under competing strength requirements}

\author[inst1,inst2,inst4]{Ashwini Gupta}

\affiliation[inst1]{organization={Department of Mechanical Engineering, Johns Hopkins University},
            city={Baltimore},
            state={Maryland},
            postcode={21218}, 
            country={USA}}

\affiliation[inst2]{organization={Department of Civil and Systems Engineering, Johns Hopkins University},
            city={Baltimore},
            state={Maryland},
            postcode={21218}, 
            country={USA}}

\affiliation[inst3]{organization={Department of Materials Science and Engineering, Johns Hopkins University},
            city={Baltimore},
            state={Maryland},
            postcode={21218}, 
            country={USA}}

\affiliation[inst4]{organization={Hopkins Extreme Materials Institute, Johns Hopkins University},
            city={Baltimore},
            state={Maryland},
            postcode={21218}, 
            country={USA}}
            
\author[inst2]{Indrashish Saha}

\author[inst1,inst2,inst3,inst4]{Lori Graham-Brady}

\author[inst1,inst4]{Tamer A. Zaki}

\begin{abstract}
Computational design of polycrystalline microstructures for enhanced mechanical performance requires repeated high-fidelity simulations over stochastic morphologies, rendering conventional crystal plasticity finite element \\
(CPFE) approaches prohibitively expensive for optimization. Here, we develop a deep learning–driven framework for optimizing polycrystalline microstructures under competing quasi-static and dynamic performance requirements. 
A 3D U-Net surrogate maps polycrystalline copper microstructures directly to full-field velocity histories from plate-impact simulations, preserving the spatial and temporal resolution needed to evaluate dynamic performance and interrogate underlying wave interactions and failure mechanisms.
The surrogate is coupled with stochastic microstructure generation and derivative-free optimization to efficiently navigate the design space.  
To monitor surrogate reliability during optimization, we introduce a distribution-consistent reliability region (DCRR) based on maximum mean discrepancy (MMD) to quantify distributional shift from the training data.
The framework is demonstrated for multiple design problems, including constrained optimization of spall and yield strengths and inverse design. 
Across the constrained optimization cases, the optimal designs achieve up to a 29\% increase in spall strength while satisfying the prescribed yield strength constraint.
The optimal designs are validated using CPFE simulations with less than 3$\%$ prediction error across all optimization cases, while the surrogate-driven optimization reduces cumulative evaluation time by approximately four orders of magnitude compared with direct CPFE-based optimization. Although demonstrated here for the design of polycrystalline microstructures targeting spall and yield strengths, the framework is broadly applicable to optimization problems where high-fidelity full-field simulations are prohibitively expensive but spatially and temporally resolved response information is essential for evaluating, interpreting, and optimizing candidate designs. 
\end{abstract}


\end{frontmatter}

\section{Introduction}

Metal alloys play a central role in high-performance structural applications across aerospace, defense, and transportation. Their performance under quasi-static and dynamic loading directly affects the safety and durability of engineered systems. Under quasi-static loading, yield strength characterizes resistance to permanent deformation, whereas under high-strain-rate loading, spall strength characterizes resistance to dynamic tensile failure~\cite{antoun2003spall}. Because these properties depend differently on the underlying microstructure, designing microstructures to simultaneously satisfy both performance requirements presents a challenging materials design problem~\cite{mcdowell2008concurrent}.

The design of polycrystalline microstructures is inherently complex because variations in grain size, morphology, orientation, and spatial arrangement can influence multiple macroscale mechanical properties~\cite{mallick2021spall, ragan2026microstructural, minich2004effect, wilkerson2016unraveling, escobedo2011effects, fromm2009grain, abdolvand2018strong, haouala2018analysis}. Consequently, explicitly incorporating microstructural information into computational materials design can be important for effectively navigating process-structure-property relationships~\cite{molkeri2022importance}.
Moreover, a prescribed set of microstructural descriptors does not define a unique microstructure, but rather an ensemble of stochastic realizations. Computational design therefore requires exploring variations in the underlying descriptors while also accounting for the variability among microstructures associated with each candidate design~\cite{generale2024inverse}. This challenge is particularly important under dynamic loading conditions, where wave propagation, grain-scale interactions, and localization influence the resulting failure response.

Traditionally, the design of high-performance alloys has relied on empirical knowledge, experiments, and physics-based simulations such as molecular dynamics (MD) and finite element (FE) methods~\cite{lieberman2016microstructural, vogler2008heterogeneous, remington2018spall, ragan2026microstructural}. 
Recent advances in automated and high-throughput experimentation are expanding the availability of data for characterizing materials under extreme loading conditions~\cite{hufnagel2026aimd}.
Wanchoo et al.~\cite{wanchoo2026high} developed an automated laser-driven microflyer impact platform capable of rapidly generating large spall datasets, creating new opportunities for data-driven investigation and materials design.
Crystal plasticity finite element (CPFE) modeling complements such experimental capabilities by providing a physics-based approach for resolving the relationship between microstructure and mechanical response. However, CPFE models of spall, which resolve grain-scale deformation and failure mechanisms under dynamic loading, are computationally expensive. Direct optimization would require repeated simulations across both candidate designs and stochastic microstructure realizations, making microstructure optimization using CPFE alone prohibitively expensive.

Recent advances in artificial intelligence and machine learning have enabled data-driven modeling of complex microstructure-property relationships, offering a promising pathway for accelerating multiscale materials design~\cite{barmak2026machine, liu2015machine, oishi2017computational, bock2019review, yang2019establishing, bhaduri2022stress, gupta2023accelerated, chen2025predicting}. 
Different approaches have demonstrated the ability to learn spatial statistics of complex microstructures and generate diverse synthetic realizations~\cite{buzzy2026polymicros, bhaduri2021efficient}.
Applications include high-throughput alloy development, process optimization, microstructure-informed property prediction, multiscale materials optimization, and inverse materials design~\cite{gao2025recent,zuo2026strength,qin2026multidimensional,kovachki2022multiscale,attari2023towards}.
Frameworks such as materials knowledge systems have similarly enabled efficient surrogate modeling using reduced microstructural representations~\cite{kalidindi2020feature}.
However, many existing approaches use reduced microstructural representations to target a limited set of scalar properties~\cite{toda2023surrogate,liu2015predictive}.
While computationally efficient, these surrogates do not directly retain the spatially resolved response through which microstructure influences dynamic failure. For problems such as spall, retaining the full-field response offers an additional advantage: the surrogate can provide both an efficient performance prediction and a means of interrogating how grain morphology alters wave interactions and failure mechanisms.

A second challenge arises when data-driven surrogates are embedded within optimization. Unlike forward prediction, where inputs may be sampled from a prescribed distribution, optimization deliberately searches for increasingly favorable and potentially unusual regions of design space. Candidate optima may therefore progressively depart from the distribution represented in the surrogate training data, precisely where prediction errors can become difficult to assess. This challenge is closely related to out-of-distribution (OOD) detection in machine learning, where distributional similarity metrics can be used to identify inputs for which model predictions may be unreliable
~\cite{hendrycks2016baseline, rabanser2019failing}. A robust surrogate-based design framework therefore requires not only computational efficiency, but also a means of monitoring whether candidate designs remain sufficiently supported by the training data~\cite{perdikaris2017nonlinear}. Maximum mean discrepancy (MMD) provides a nonparametric measure of the discrepancy between distributions, offering a natural basis for quantifying how candidate microstructure ensembles depart from the distribution represented during surrogate training~\cite{gretton2012kernel}.

\begin{figure}[htbp!]
\centering
\includegraphics[width=0.90\columnwidth,keepaspectratio]{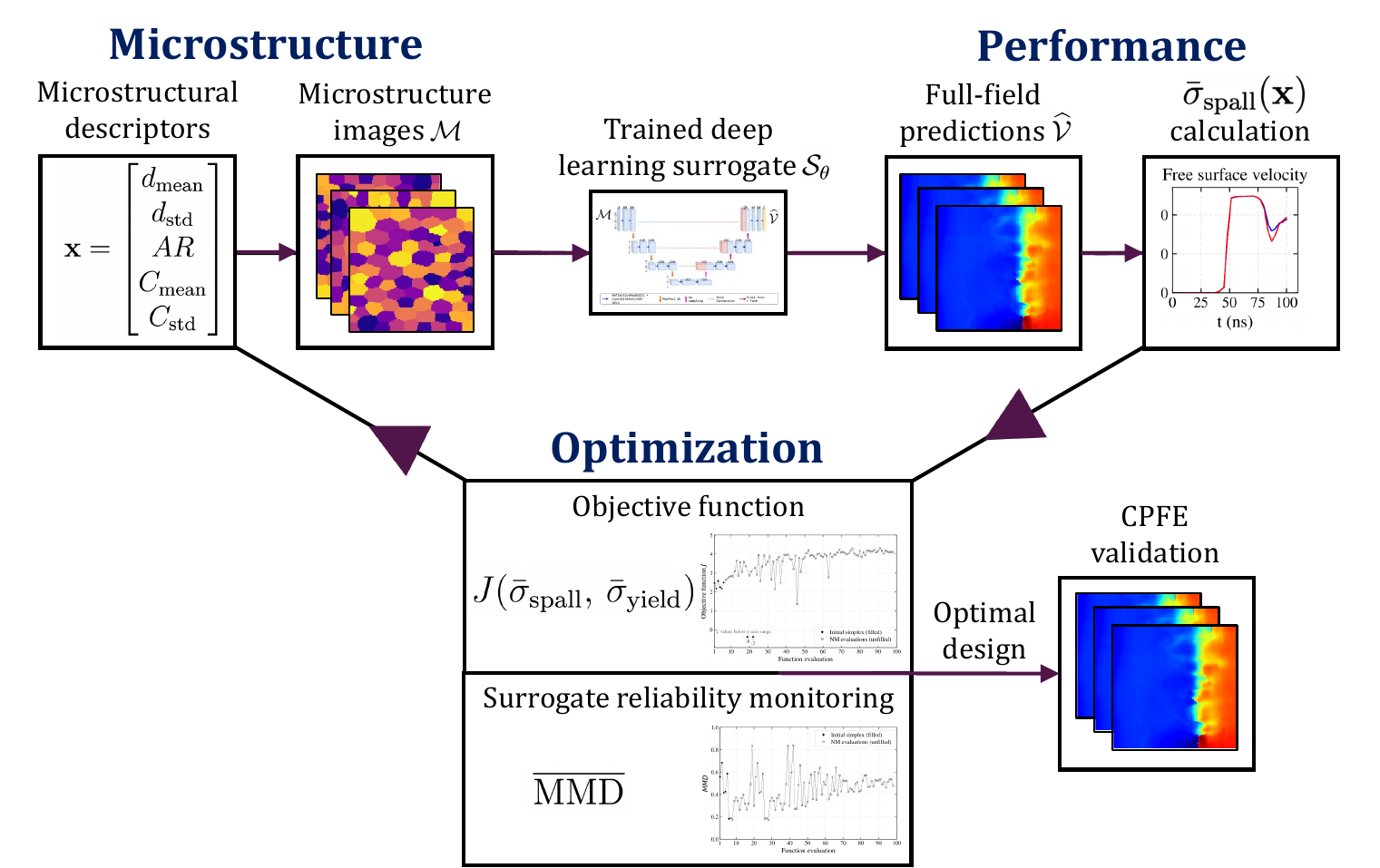}
\caption{Overview of the deep learning-driven optimization framework. Microstructural descriptors define candidate designs, from which multiple stochastic microstructure realizations are generated. The trained surrogate predicts the corresponding full-field velocity histories, from which the free-surface velocity history is extracted to calculate the mean spall strength. The optimizer iteratively updates the microstructural descriptors, while MMD quantifies distributional shift and monitors surrogate reliability during optimization. The optimal designs are validated using high-fidelity CPFE simulations.
}
\label{fig:optimization_overview}
\end{figure}

This work presents a deep learning–driven framework for reliability-aware optimization of polycrystalline microstructures under competing quasi-static and dynamic performance requirements. The framework makes three primary contributions. First, a 3D U-Net surrogate maps polycrystalline microstructures to full-field velocity histories from plate-impact simulations, retaining spatially and temporally resolved response information that supports both efficient evaluation of spall strength and mechanistic interpretation of dynamic response. Second, the surrogate is coupled with stochastic microstructure generation and derivative-free optimization to address multiple design problems, including constrained optimization of spall and yield strengths as well as inverse design for target properties. Third, a distribution-consistent reliability region (DCRR), based on maximum mean discrepancy (MMD), is introduced to quantify distributional shift and monitor surrogate reliability as optimization navigates the microstructural design space. The resulting optimal designs are validated against high-fidelity CPFE simulations. An overview of the proposed framework is shown in Fig.~\ref{fig:optimization_overview}.

The paper is organized as follows. Section~\ref{sec:optimization_problem} formulates the optimization problem and presents an overview of the proposed surrogate-driven framework. Section~\ref{sec:methodology} describes microstructure generation, property evaluation, the deep learning model, and the distribution-aware surrogate reliability framework. Section~\ref{sec:results} presents the optimization results across different design scenarios, along with mechanistic interpretation of the resulting microstructure–property relationships and an analysis of computational efficiency. Finally, Section~\ref{sec:conclusions} summarizes the main findings and outlines directions for future work.

\section{Optimization problem formulation}\label{sec:optimization_problem}
The microstructure design problem is formulated as a set of nonlinear optimization problems over a bounded admissible design space $\Omega$. Microstructural descriptors characterizing grain morphology are employed as design variables. The five-dimensional design vector $\mathbf{x} \in \mathbb{R}^5$ is defined as
\begin{equation}\label{eq:design_vars}
\mathbf{x} =
\left(
d_{\mathrm{mean}},\;
d_{\mathrm{std}},\;
AR,\;
C_{\mathrm{mean}},\;
C_{\mathrm{std}}
\right),
\end{equation}
where the design variables $d_{\mathrm{mean}}$ and $d_{\mathrm{std}}$ denote the mean and standard deviation of the grain size distribution, respectively; $AR$ represents the average grain aspect ratio; and $C_{\mathrm{mean}}$ and $C_{\mathrm{std}}$ denote the mean and standard deviation of grain circularity, respectively.

The design variables are subject to bound constraints,
\begin{equation}
x_i^{\min} \le x_i \le x_i^{\max},
\qquad i = 1, \ldots, 5,
\end{equation}
which define the admissible design space $\Omega \subset \mathbb{R}^5$. The bounds for each variable are summarized in Table~\ref{tab:design_bounds}. These constraints define the range of microstructural descriptors considered in the optimization and restrict the search to a prescribed region of the design space.

\begin{table}[htbp]
\centering
\caption{Bounds on microstructural design variables.}
\label{tab:design_bounds}
\begin{tabular}{l
                S[table-format=2.3]
                S[table-format=2.3]}
\hline
Variable & {Lower bound} & {Upper bound} \\
\hline
$d_{\mathrm{mean}}$ ($\mu\mathrm{m}$) & 18.00 & 65.00 \\
$d_{\mathrm{std}}$ ($\mu\mathrm{m}$)  &  2.00 & 10.00 \\
$AR$                                  &  0.40 &  2.50 \\
$C_{\mathrm{mean}}$                   &  0.94 &  0.98 \\
$C_{\mathrm{std}}$                    &  0.02 &  0.06 \\
\hline
\end{tabular}
\end{table}

Because each design vector $\mathbf{x}$ defines an ensemble of stochastic microstructures, the optimization must account for variability among microstructure realizations associated with a given design. Therefore, given a design vector $\mathbf{x}$, the stochastic microstructure generator produces realizations $\mathcal{M}(\mathbf{x}, \xi_s)$, where variability is governed by the random seed $\xi_s$. The spall strength $\sigma_{\mathrm{spall}}$ is evaluated for each realization and thus exhibits variability even for a fixed $\mathbf{x}$. Accordingly, the mean spall strength $\bar{\sigma}_{\mathrm{spall}}(\mathbf{x})$ is estimated via Monte Carlo sampling:
\begin{equation}
\bar{\sigma}_{\mathrm{spall}}(\mathbf{x})
=
\frac{1}{N_{\mathrm{micro}}}\sum_{s=1}^{N_{\mathrm{micro}}} \sigma_{\mathrm{spall}}\big(\mathcal{M}(\mathbf{x}, \xi_s)\big),
\qquad N_{\mathrm{micro}} = 100.
\label{eq:spall_mc_avg}
\end{equation}
Hereafter, when referring to a microstructural design, “spall strength” denotes $\bar{\sigma}_{\mathrm{spall}}(\mathbf{x})$.

The proposed framework supports two classes of microstructure design problems: (i) constrained maximization of spall strength under a yield strength constraint, and (ii) target-matching inverse design.

(i) For constrained maximization, the objective is to maximize mean spall strength while maintaining a prescribed minimum yield strength $\bar{\sigma}_{\mathrm{yield}}^{\mathrm{min}}$. This is posed as
\begin{equation}\label{eq:constrained_problem}
\max_{\mathbf{x} \in \Omega} 
\quad 
\bar{\sigma}_{\mathrm{spall}}(\mathbf{x})
\quad \text{s.t.} \quad
\bar{\sigma}_{\mathrm{yield}}(\mathbf{x})
\ge 
\bar{\sigma}_{\mathrm{yield}}^{\mathrm{min}},
\end{equation}
where $\bar{\sigma}_{\mathrm{yield}}(\mathbf{x})$ denotes the mean yield strength. 

(ii) For target-matching inverse design, the objective is to minimize the mismatch between the mean spall and yield strength of a microstructural design and their prescribed target values $\bar{\sigma}_{\mathrm{spall}}^{\mathrm{target}}$ and $\bar{\sigma}_{\mathrm{yield}}^{\mathrm{target}}$, respectively. This is posed as
\begin{equation}\label{eq:target_matching}
\min_{\mathbf{x} \in \Omega}
\quad
\operatorname{dist}
\left(
\begin{bmatrix}
\bar{\sigma}_{\mathrm{spall}}(\mathbf{x}) \\
\bar{\sigma}_{\mathrm{yield}}(\mathbf{x})
\end{bmatrix},
\begin{bmatrix}
\bar{\sigma}_{\mathrm{spall}}^{\mathrm{target}} \\
\bar{\sigma}_{\mathrm{yield}}^{\mathrm{target}}
\end{bmatrix}
\right),
\end{equation}
where $\operatorname{dist}(\cdot,\cdot)$ denotes a measure of the mismatch between the design and target property values. The numerical solution of both optimization formulations is described in the following section.

\section{Methodology}\label{sec:methodology}
An overview of the proposed surrogate-driven optimization framework is shown in Fig.~\ref{fig:optimization_overview}. The framework integrates stochastic microstructure generation, surrogate-based evaluation, and derivative-free optimization to enable efficient microstructure design. Microstructures are parameterized using statistical descriptors that characterize stochastic grain morphologies, from which multiple realizations are generated to capture stochastic variability. For each candidate design, the trained deep learning surrogate predicts the corresponding full-field velocity histories, from which the mean spall strength is evaluated. The yield strength is computed from the mean grain size using the Hall–Petch relation. The constrained optimization problem is solved using the derivative-free Nelder–Mead algorithm, which iteratively updates the microstructural descriptors based on surrogate evaluations. During optimization, maximum mean discrepancy (MMD) between the latent representations of candidate and training microstructures quantifies distributional shift and serves as a monitoring tool for surrogate reliability. The surrogate predictions for the optimal design are subsequently validated against high-fidelity CPFE simulations. The individual components of the framework are detailed in the following subsections.

\subsection{Stochastic microstructure generation}\label{sec:micro_generation}
A stochastic microstructure generation procedure is employed to construct polycrystalline microstructures from the five microstructural descriptors defined in Eq.~\eqref{eq:design_vars}. These descriptors parameterize the grain morphology and define the design space considered in the optimization. Polycrystalline microstructures are generated using the open-source tessellation software \textit{Neper} \cite{quey2011large}. For a given 
design vector $\mathbf{x}$ and random seed $\xi_s$, Neper generates a 
two-dimensional microstructure realization $\mathcal{M}(\mathbf{x},\xi_s)$. This stochastic generation procedure provides the microstructure ensembles required for training the surrogate model and evaluating candidate designs during optimization.

Neper constructs two-dimensional Laguerre-Voronoi tessellations within a \(200\times200~\mu\mathrm{m}^2\) square domain. Grain size is prescribed by a lognormal distribution of equivalent grain diameter with mean \(d_{\mathrm{mean}}\) and standard deviation \(d_{\mathrm{std}}\), whereas grain shape is controlled by the aspect ratio \(AR\) and circularity \(C\). In Neper, circularity is internally parameterized through a lognormal distribution of \(1-C\); all quantities reported in this work are expressed in terms of \(C_{\mathrm{mean}}\) and \(C_{\mathrm{std}}\).

To construct the microstructure inputs for the training dataset $\mathcal{D}_{\mathrm{train}}$, descriptor combinations are sampled throughout the admissible design space $\Omega$ defined by the bounds in Table~\ref{tab:design_bounds}. Latin hypercube sampling is employed to provide space-filling coverage of the design space, subject to the additional constraint $d_{\mathrm{std}} \leq d_{\mathrm{mean}}/2$. The prescribed bounds and constraint limit the sampling to the range of microstructural descriptors considered in this work and reduce the generation of non-physical or geometrically inconsistent microstructures. Additional bounds or constraints could be incorporated to account for manufacturing limitations when such information is available.

\subsection{Property evaluation}\label{sec:property_evaluation}
Two material properties are evaluated for each microstructure design: the mean spall strength $\bar{\sigma}_{\mathrm{spall}}$ and the mean yield strength $\bar{\sigma}_{\mathrm{yield}}$. Their evaluation is described in the following subsections.

\subsubsection{Spall strength calculation}\label{sec:spall_calculation}
Spall is a dynamic tensile failure process in which interacting stress waves generate sufficiently large tensile stresses within a material to produce internal damage and fracture. Here, spall strength is evaluated using crystal plasticity finite element (CPFE) simulations of plate-impact loading in polycrystalline copper at loading strain rates of approximately $O(10^6~\mathrm{s}^{-1})$. The numerical framework follows Saha and Graham-Brady \cite{saha2026numerical} and couples crystal plasticity with a cohesive-zone formulation to capture grain-scale deformation and intergranular failure during spall. The loading configuration, constitutive parameters, and damage model are adopted from \cite{saha2026numerical}. All geometric, loading, and material parameters are held fixed across microstructures, such that variations in spall response arise from differences in grain morphology. The complex, microstructure-dependent physics of spall, coupled with the high computational cost of CPFE simulations, makes it a compelling test bed for the proposed surrogate-based optimization framework.

For a given microstructure $\mathcal{M}(\mathbf{x},\xi_s)$, a CPFE simulation is performed to model stress-wave propagation and spall failure. The simulation produces a microstructure-resolved full-field velocity history $\mathcal{V}\!\left(\mathcal{M}(\mathbf{x},\xi_s)\right)$, hereafter denoted simply by $\mathcal{V}$, from which the spatially averaged free-surface velocity history $V_{fs}$ is extracted. The Neper-generated microstructures and their corresponding CPFE-computed full-field velocity histories constitute the input-output pairs of the training dataset $\mathcal{D}_{\mathrm{train}}$ for the deep learning surrogate, as described in Section~\ref{sec:deep_learning_model}. Once trained, the surrogate replaces repeated CPFE evaluations during optimization and predicts the corresponding full-field velocity history, denoted by $\widehat{\mathcal{V}}$.

\begin{figure*}[htbp]
\centering
\begin{subfigure}{0.7\columnwidth}
    \centering
    \includegraphics[width=\linewidth]{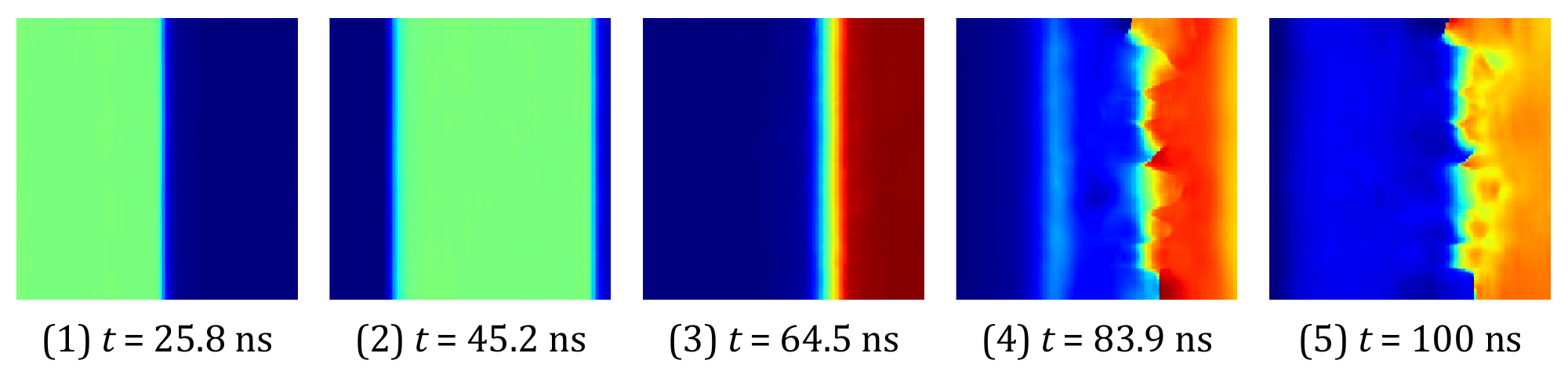}
    \caption{}
    \label{fig:vel_pred_unet}
\end{subfigure}
\par
\vspace{4mm}
\begin{subfigure}{0.5\columnwidth}
    \centering
    \includegraphics[width=\linewidth]{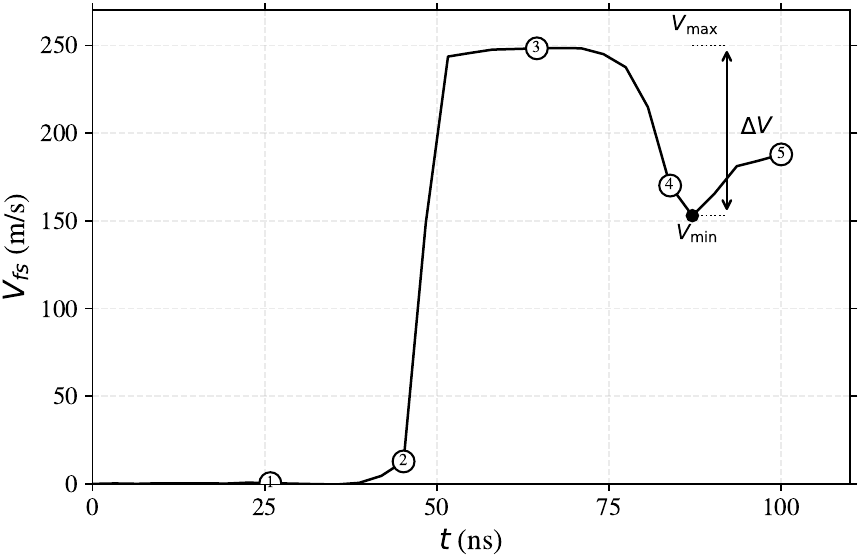}
    \caption{}
    \label{fig:vfs_calc_unet}
\end{subfigure}
\caption{Extraction of spall strength from the surrogate-predicted dynamic response. (a) Representative snapshots of the surrogate-predicted full-field velocity history $\widehat{\mathcal{V}}$ at selected time points. (b) Corresponding free-surface velocity history $V_{fs}$, where markers 1-5 indicate the times associated with the snapshots in (a), and $\Delta V$ denotes the pull-back velocity used to calculate the spall strength.}
\label{fig:spall_extraction}
\end{figure*}

The same procedure is used to extract spall strength from either the CPFE-computed or surrogate-predicted full-field response. Figure~\ref{fig:spall_extraction} illustrates this procedure using a representative surrogate-predicted full-field velocity history.
Under shock loading, the interaction of release waves produces a characteristic pull-back in $V_{fs}$ associated with spall formation. The spall strength $\sigma_{\mathrm{spall}}$ is calculated from the pull-back velocity drop $\Delta V$ as
\begin{equation}
\sigma_{\mathrm{spall}} = \frac{1}{2}\rho_0 C_l \Delta V ,
\end{equation}
where $\rho_0$ is the material density, $C_l$ is the longitudinal wave speed, and $\Delta V$ is the difference between the peak velocity and the subsequent pull-back minimum~\cite{antoun2006spall}. For polycrystalline copper, $\rho_0 = 8960 ~\mathrm{kg/m^3}$ and $C_l = 3933 ~\mathrm{m/s}$ are used throughout this work.

\begin{figure*}[htbp]
\centering
\begin{subfigure}{0.49\textwidth}
    \centering
    \includegraphics[width=0.6\linewidth]{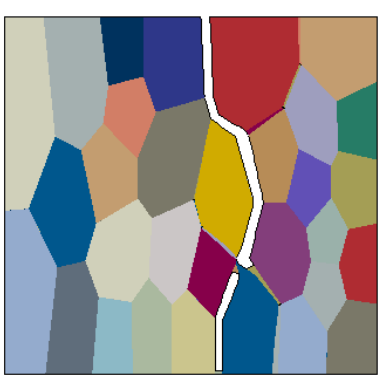}
    \caption{}
    \label{fig:crack_columnar}
\end{subfigure}
\hfill
\begin{subfigure}{0.49\textwidth}
    \centering
    \includegraphics[width=0.6\linewidth]{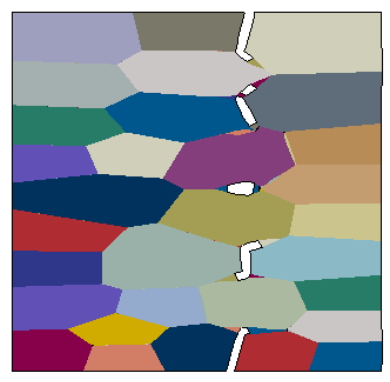}
    \caption{}
    \label{fig:crack_elongated}
\end{subfigure}
\caption{Effect of grain morphology on interfacial separation and spall-plane formation. (a) Columnar grains promote predominantly mode-I (tensile) crack opening along grain boundaries. (b) Elongated grains promote mixed-mode (tensile + shear) crack opening along tortuous grain-boundaries.}
\label{fig:crack_modes}
\end{figure*}

Figure~\ref{fig:crack_modes} illustrates the influence of grain morphology on interfacial separation and spall-plane formation. Columnar grains favor predominantly mode-I (tensile) crack opening along grain boundaries, leading to relatively planar spall paths, whereas elongated grains promote mixed-mode tensile and shear separation along tortuous grain boundaries, resulting in more diffuse and irregular spall-plane formation. These observations underscore the role of grain morphology in governing interfacial failure modes and, consequently, the resulting spall strength. This morphology dependence provides physical motivation for optimizing the microstructural descriptors to enhance spall performance.

During optimization, for each microstructure design $\mathbf{x}$, $N_{\mathrm{micro}}=100$ stochastic realizations are generated using distinct $\xi_s$ and evaluated using the deep learning surrogate. The corresponding spall strengths are averaged according to Eq.~\eqref{eq:spall_mc_avg} to obtain $\bar{\sigma}_{\mathrm{spall}}$. This averaging accounts for stochastic variability in grain morphology among realizations of a fixed design within the prescribed simulation domain.

\subsubsection{Yield strength calculation}
The mean yield strength is estimated using the Hall-Petch relationship, which relates the macroscopic yield strength to the mean grain size of the microstructure. Because the Hall-Petch estimate depends only on $d_{\mathrm{mean}}$, $\bar{\sigma}_{\mathrm{yield}}$ is invariant across stochastic realizations generated for a fixed design vector $\mathbf{x}$. The mean yield strength used in the optimization is therefore given by
\begin{equation}
\bar{\sigma}_{\mathrm{yield}} = \sigma_0 + k\, d_{\mathrm{mean}}^{-1/2},
\end{equation}
where $\sigma_0$ is the lattice friction stress, $k$ is the Hall-Petch coefficient for polycrystalline copper, and $d_{\mathrm{mean}}$ is the mean grain size of the microstructure. For polycrystalline copper, $\sigma_0 = 25~\mathrm{MPa}$ and $k = 0.11~\mathrm{MPa}\sqrt{\mathrm{m}}$ are used throughout this work.
Within the optimization framework, $\bar{\sigma}_{\mathrm{yield}}$ is used to impose the prescribed yield strength constraint or target, depending on the design problem.

\subsection{Deep learning model}\label{sec:deep_learning_model}
To avoid repeated high-fidelity CPFE simulations during optimization, a deep learning model is employed as a surrogate for the microstructure-dependent dynamic response. The surrogate $\mathcal{S}_{\theta}$ approximates the mapping from a microstructure $\mathcal{M}$ to its corresponding full-field velocity history $\mathcal{V}$,
\begin{equation}
\widehat{\mathcal{V}}=\mathcal{S}_{\theta}(\mathcal{M}),
\end{equation}
where $\widehat{\mathcal{V}}$ denotes the surrogate-predicted full-field velocity history and $\theta$ represents the learned model parameters. 
As illustrated in Fig.~\ref{fig:spall_extraction}, $\widehat{\mathcal{V}}$ is subsequently used to extract the free-surface velocity history and evaluate spall strength.
By replacing repeated CPFE simulations with rapid surrogate evaluations, the model enables efficient optimization over the microstructural design space.

\subsubsection{Model architecture and training}
The surrogate employs a 3D U-Net architecture previously developed in~\cite{saha2026numerical, saha2024prediction}. The network maps 2D microstructure images to their corresponding full-field velocity histories. The encoder-decoder architecture with skip connections enables multiscale feature extraction while retaining spatial and temporal information in the predicted response. The network consists of three encoder levels, beginning with 64 feature channels that double at each subsequent level. Convolutional layers employ \(5\times5\times5\) kernels, followed by group normalization and GELU activation. Further architectural details are provided in~\cite{saha2026numerical, saha2024prediction}. The training data $\mathcal{D}_{\mathrm{train}}$ consist of microstructure-response pairs generated using CPFE simulations. Each microstructure and its corresponding velocity fields are spatially discretized on a uniform $128\times128$ grid. In the temporal dimension, 32 frames are uniformly sampled over the interval \(t=0\) to \(100\) ns, resulting in a \(128\times128\times32\) spatiotemporal representation of the velocity response. $\mathcal{D}_{\mathrm{train}}$ comprises 640 microstructure realizations, while 80 additional realizations are used as a validation set for model selection.

Training is performed by minimizing the mean squared error (MSE) between the surrogate-predicted and CPFE-computed full-field velocity histories. Model parameters are optimized using the Adam optimizer with an initial learning rate of $10^{-3}$, which is adaptively reduced based on the validation loss. Mini-batch training is performed with a batch size of $N_b=32$ for up to $N_{\mathrm{epochs}}=200$ epochs, with early stopping based on validation performance. The final model corresponds to the epoch with the lowest validation loss. Training on a single NVIDIA A100 GPU requires approximately 8 hours.

\subsubsection{Model performance and generalization}
To evaluate predictive accuracy and generalization capability, a separate test dataset $\mathcal{D}_{\mathrm{test}}$ comprising 1140 microstructure realizations is generated exclusively for independent model evaluation. This dataset is not used during training or hyperparameter tuning and spans a broad range of microstructural descriptors, including regions within and beyond the ranges represented in the training data.

\begin{figure*}[htbp]
\centering
\begin{subfigure}{0.75\columnwidth}
    \centering
    \includegraphics[
    width=\linewidth,
    trim=0 0.5cm 0 1.5cm,  
    clip
    ]{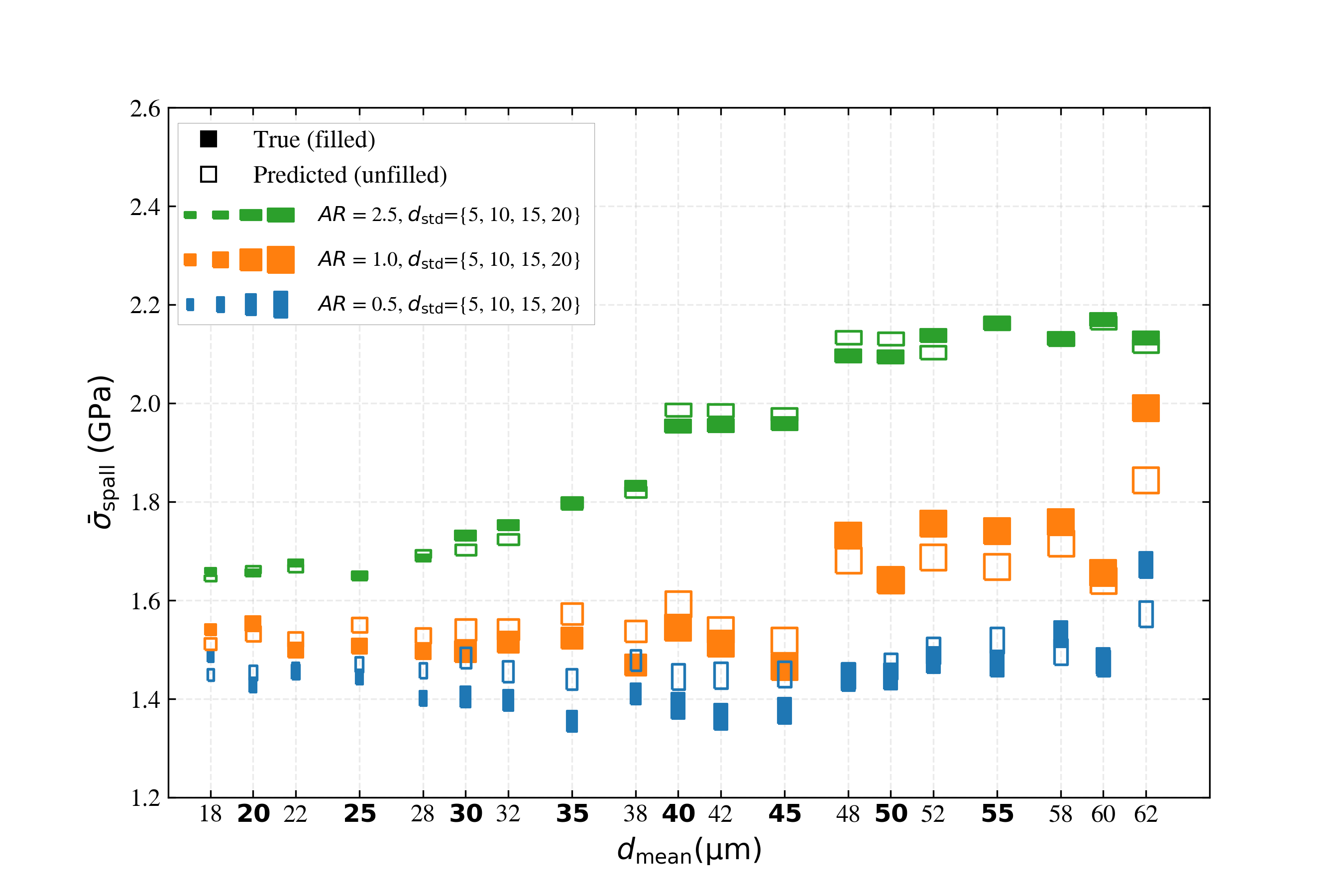}
    \caption{}
    \label{fig:PRTR_vs_grainSize}
\end{subfigure}
\par
\vspace{2mm}
\begin{subfigure}{0.75\columnwidth}
    \centering
    \includegraphics[
    width=\linewidth,
    trim=0 0.5cm 0 1.5cm,
    clip
    ]{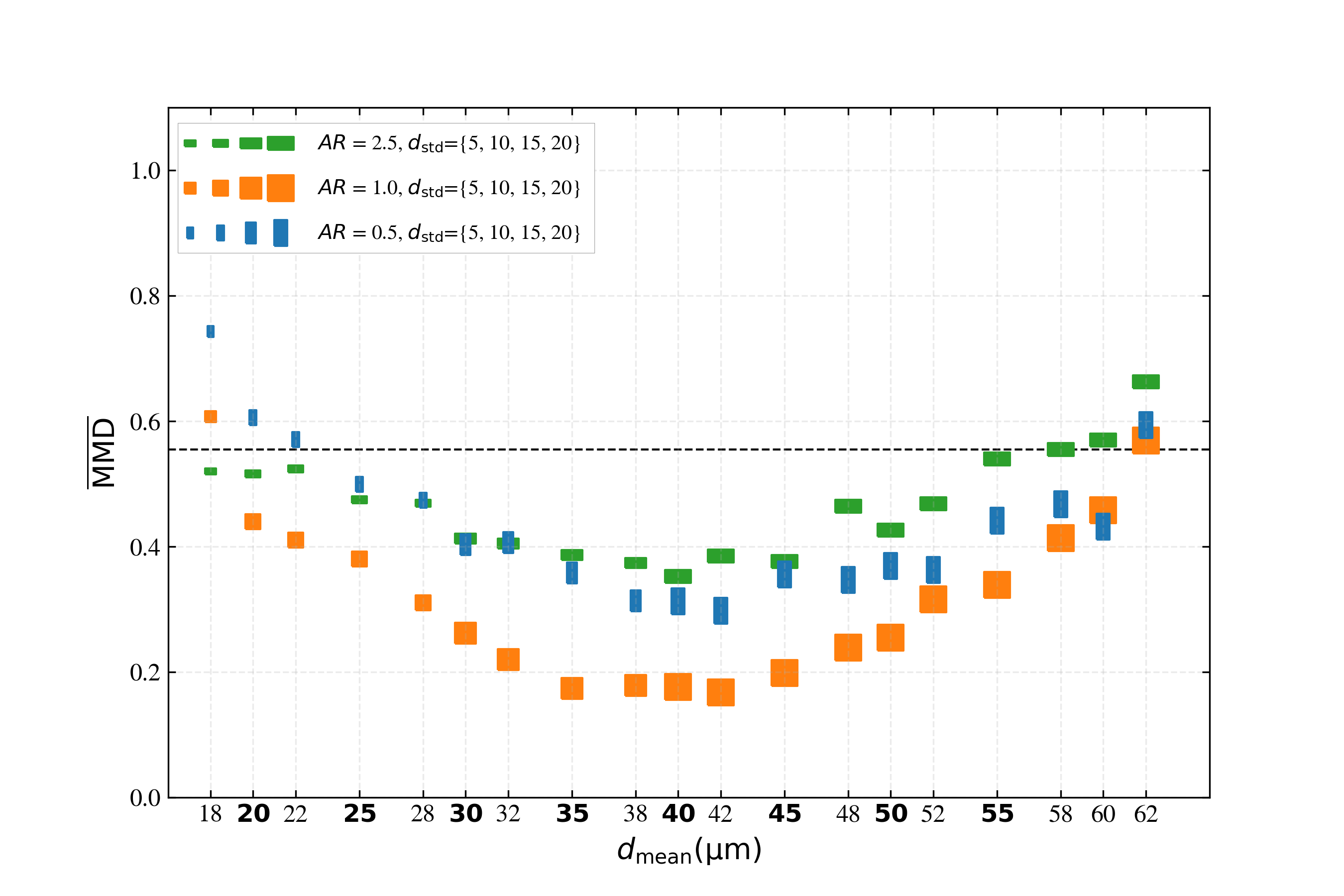}
    \caption{}
    \label{fig:MMD_vs_grainSize}
\end{subfigure}
\caption{Surrogate performance and distributional shift for different combinations of $d_{\mathrm{mean}}$, $d_{\mathrm{std}}$, and $AR$: (a) CPFE-computed reference and surrogate-predicted $\bar{\sigma}_{\mathrm{spall}}$ and (b) $\overline{\mathrm{MMD}}$. Bold tick labels denote $d_{\mathrm{mean}}$ values represented in the training set, highlighting interpolation and extrapolation with respect to mean grain size. The horizontal dashed line in (b) denotes the DCRR threshold $\tau=0.55$.}\label{fig:model_performance}
\end{figure*}

Model performance is assessed by comparing surrogate-predicted responses with the corresponding CPFE-computed results. The relative error of a quantity $X$ with respect to a reference value $X_{\mathrm{ref}}$ is defined as
\begin{equation}\label{eq:percent_relative_error}
\varepsilon_X\,(\%) = 100\,\frac{|X - X_{\mathrm{ref}}|}{X_{\mathrm{ref}}}.
\end{equation}
This dimensionless metric quantifies the deviation of the surrogate prediction from the reference CPFE-computed value.

Figure~\ref{fig:PRTR_vs_grainSize} shows the surrogate-predicted and CPFE-computed $\bar{\sigma}_{\mathrm{spall}}$ for microstructures in test dataset with varying $d_{\mathrm{mean}}$, $d_{\mathrm{std}}$, and $AR$. All cases use fixed circularity descriptors with $C_\mathrm{mean} = 95\,\%$ and $C_\mathrm{std} = 2\,\%$. Each point here represents $\bar{\sigma}_{\mathrm{spall}}$ averaged over 20 statistically equivalent microstructures. Filled markers denote CPFE-computed values, while unfilled markers denote the corresponding surrogate predictions. Bold tick labels denote the mean grain sizes used in the training set, highlighting interpolation and extrapolation with respect to mean grain size. The corresponding absolute relative error $\varepsilon_{\bar{\sigma}_{\mathrm{spall}}}\,(\%)$ remains below 8$\%$ across all cases.

Overall, the surrogate predictions closely follow the CPFE-computed trends across the parameter space, with the largest deviations observed at lower aspect ratios. The model maintains low prediction errors across the evaluated descriptor ranges, demonstrating generalization to both interpolated and extrapolated microstructures. Distributional shift between candidate microstructures and the training distribution is quantified using MMD, as described in the following section.

\subsection{Distribution-aware surrogate reliability framework}\label{sec:reliability_framework}

Surrogate-driven optimization can lead to extrapolative evaluations when candidate designs depart from the distribution represented by the training data. Maximum mean discrepancy (MMD), a kernel-based two-sample statistic widely used for comparing probability distributions~\cite{gretton2012kernel}, is employed here to quantify distributional shift between candidate microstructures and the training distribution. Based on this discrepancy, we define a distribution-consistent reliability region (DCRR) to monitor surrogate reliability during optimization.

To obtain a learned feature representation suitable for distributional comparison, features are extracted from an intermediate encoder layer of the trained 3D U-Net surrogate. 
Specifically, the encoder layer \path{enc3.conv\_block.conv2} is selected as it provides a deep representation of morphology-relevant and prediction-relevant features for which MMD exhibits a strong correspondence with surrogate prediction error.
Let $\phi(\cdot)$ denote the feature mapping defined by this layer. 
For a microstructure $\mathcal{M}(\mathbf{x},\xi_s)$, the corresponding feature embedding is defined as
\begin{equation}
\mathbf{z} = \phi\!\left(\mathcal{M}(\mathbf{x},\xi_s)\right) \in \mathbb{R}^{d_z},
\end{equation}
where $d_z$ denotes the dimensionality of the flattened feature tensor at this layer. 
The training dataset consists of $n=640$ microstructure realizations, whose corresponding feature embeddings define the reference feature set
\begin{equation*}
\mathcal{Z}_{\mathrm{train}} = \{ \mathbf{z}_i \}_{i=1}^{n},
\end{equation*}
where each $\mathbf{z}_i$ is extracted from the same encoder layer.

MMD is evaluated in this learned feature space rather than in the raw five-dimensional descriptor space. Euclidean proximity in descriptor space does not necessarily reflect similarity in higher-order morphological features relevant to surrogate predictions. In contrast, the encoder activations provide a learned representation of the microstructure that captures spatial and morphological features associated with the predicted response, making the feature space more informative for quantifying distributional shift relative to the training data.

\subsubsection{Quantifying distributional shift via maximum mean discrepancy}\label{sec:MMD}

MMD is used to quantify distributional shift between candidate microstructures and the training distribution in the learned feature space. For two distributions $P$ and $Q$ in feature space, the squared MMD is defined as 
\begin{equation}
\mathrm{MMD}^2(P,Q;k)
=
\mathbb{E}_{\mathbf{z},\mathbf{z}'\sim P}[k(\mathbf{z},\mathbf{z}')]
+
\mathbb{E}_{\mathbf{w},\mathbf{w}'\sim Q}[k(\mathbf{w},\mathbf{w}')]
-
2\mathbb{E}_{\mathbf{z}\sim P,\mathbf{w}\sim Q}[k(\mathbf{z},\mathbf{w})],
\end{equation}
where $\mathbf{z}$ and $\mathbf{z}'$ are feature vectors sampled from $P$, $\mathbf{w}$ and $\mathbf{w}'$ are feature vectors sampled from $Q$, and $k(\cdot,\cdot)$ is a positive-definite kernel. A Gaussian radial basis function (RBF) kernel is employed,
\begin{equation}
k(\mathbf{a},\mathbf{b})
=
\exp\!\left(
-\frac{\|\mathbf{a}-\mathbf{b}\|_2^2}{2\sigma^2}
\right),
\end{equation}
where $\sigma$ denotes the kernel bandwidth and is determined using the median pairwise distance among the reference feature embeddings in $\mathcal{Z}_{\mathrm{train}}$.

For a microstructure realization $\mathcal{M}(\mathbf{x},\xi_s)$ with feature embedding
\begin{equation*}
\mathbf{z}_s=\phi\!\left(\mathcal{M}(\mathbf{x},\xi_s)\right),
\end{equation*}
the empirical squared MMD relative to the reference feature set $\mathcal{Z}_{\mathrm{train}}$ is
\begin{align}
\widehat{\mathrm{MMD}}^2(\mathbf{x},\xi_s)
&=
\frac{1}{n(n-1)}
\sum_{\substack{i,i'=1\ i\neq i'}}^{n}
k(\mathbf{z}_i,\mathbf{z}_{i'})
+
k(\mathbf{z}_s,\mathbf{z}_s)
-
\frac{2}{n}
\sum_{i=1}^{n}
k(\mathbf{z}_i,\mathbf{z}_s),
\label{eq:mmd_single}
\end{align}
where $\mathbf{z}_i\in\mathcal{Z}_{\mathrm{train}}$ and $n$ denotes the number of reference embeddings.

To account for stochastic variability among microstructure realizations, the squared MMD associated with each design $\mathbf{x}$ is averaged over $N_{\mathrm{micro}}=100$ realizations:
\begin{equation}
\overline{\mathrm{MMD}}(\mathbf{x})
=
\frac{1}{N_{\mathrm{micro}}}
\sum_{s=1}^{N_{\mathrm{micro}}}
\widehat{\mathrm{MMD}}^2(\mathbf{x},\xi_s),
\qquad N_{\mathrm{micro}}=100.
\label{eq:mmd_mc_avg}
\end{equation}
Hereafter, $\overline{\mathrm{MMD}}(\mathbf{x})$ denotes the MMD associated with a microstructural design $\mathbf{x}$.

\subsubsection{Distribution-consistent reliability region (DCRR)}

The distribution-consistent reliability region (DCRR) defines a subset of the admissible design space $\Omega$ in which candidate microstructures remain distributionally consistent with the training data in feature space. The region is empirically calibrated against surrogate prediction error, providing a distribution-aware indicator of surrogate reliability as optimization navigates the design space.

The reliability region $\Omega_\tau$ is defined through an MMD threshold $\tau$ as
\begin{equation}
\Omega_\tau
=
\left\{
\mathbf{x} \in \Omega :
\overline{\mathrm{MMD}}(\mathbf{x}) \le \tau
\right\}.
\end{equation}
The subset $\Omega_\tau \subset \Omega$ therefore comprises designs whose generated microstructures remain sufficiently close to the training distribution in the learned feature space according to the prescribed threshold.

Let $\mathcal{X}_{\mathrm{test}}=\{\mathbf{x}_j^{\mathrm{test}}\}_{j=1}^{N_{\mathrm{test}}}$ denote the set of design vectors represented in the held-out test dataset $\mathcal{D}_{\mathrm{test}}$, where $N_{\mathrm{test}}$ is the number of distinct test designs. The MMD threshold $\tau$ defining the reliability region $\Omega_{\tau}$ is determined as
\begin{equation}
\tau
=
\max \left\{
t \ge 0
\;:\;
\max_{\mathbf{x}\in\mathcal{X}_{\mathrm{test}} \,:\, \overline{\mathrm{MMD}}(\mathbf{x}) \le t}
\varepsilon_{\bar{\sigma}_{\mathrm{spall}}}(\mathbf{x})
< 7
\right\}.
\end{equation}
In practice, \(\tau\) is selected from the MMD values observed in the test set as the largest value for which all test designs satisfying \(\overline{\mathrm{MMD}}\le\tau\) exhibit spall-strength prediction errors below \(7\%\). For the present surrogate and test dataset, this criterion yields $\tau=0.55$. By construction, $\Omega_{\tau}$ therefore defines a performance-calibrated region associated with the prescribed accuracy tolerance on the held-out test set.

This selection of $\tau$ is further supported by the distributional trends observed across the test designs. Figure~\ref{fig:MMD_vs_grainSize} shows $\overline{\mathrm{MMD}}$ values for microstructures in test dataset with varying $d_{\mathrm{mean}}$, $d_{\mathrm{std}}$, and $AR$. 
Bold tick labels denote $d_{\mathrm{mean}}$ values represented in the training set, all of which satisfy $\overline{\mathrm{MMD}} \le 0.55$, indicating that $\Omega_{\tau}$ encompasses the variability represented in the training data.
Moreover, $\overline{\mathrm{MMD}}$ increases as the microstructural descriptors deviate from the central ranges represented in the training data, with extreme $d_{\mathrm{mean}}$ values outside the training range exhibiting larger discrepancies.
This behavior shows that $\overline{\mathrm{MMD}}$ captures distributional shift as the microstructures depart from the training distribution and supports the use of $\Omega_{\tau}$ as an empirically calibrated reliability indicator for surrogate predictions.

It is important to emphasize that $\Omega_{\tau}$ does not constitute a probabilistic uncertainty bound, nor does it enforce strict out-of-distribution rejection. Rather, it defines a performance-calibrated, distribution-consistent region within which surrogate predictions have been empirically observed to maintain acceptable accuracy. During optimization, $\overline{\mathrm{MMD}}(\mathbf{x})$ is evaluated for each candidate design; however, designs with $\overline{\mathrm{MMD}}(\mathbf{x})>\tau$ are not discarded. Instead, such designs are flagged as lying outside the DCRR $\Omega_{\tau}$ and can be prioritized for direct CPFE validation. Accordingly, the DCRR provides a distribution-aware surrogate reliability indicator without explicitly constraining the optimization trajectory.

\subsection{Optimization algorithm}\label{sec:opt_algo}

The trained surrogate model is embedded within an optimization framework to determine optimal microstructural designs. The resulting optimization problems are formulated as bound-constrained nonlinear programs and solved using the derivative-free Nelder–Mead (NM) algorithm over the admissible design space $\Omega$.

For the constrained maximization case, the inequality constraint in Eq.~\eqref{eq:constrained_problem} is incorporated into the objective function $J$ via a quadratic penalty formulation:
\begin{equation}\label{eq:max_J}
\max_{\mathbf{x} \in \Omega} \quad 
J(\mathbf{x}) 
= \bar{\sigma}_{\mathrm{spall}}(\mathbf{x})^2
- \alpha \, \max\!\left(0,\; 
\bar{\sigma}_{\mathrm{yield}}^{\mathrm{min}}
- \bar{\sigma}_{\mathrm{yield}}(\mathbf{x})
\right)^2 ,
\end{equation}
where $\alpha > 0$ is a penalty parameter controlling the penalty associated with constraint violation. The penalty term is inactive when $\bar{\sigma}_{\mathrm{yield}}(\mathbf{x}) \ge \bar{\sigma}_{\mathrm{yield}}^{\mathrm{min}}$ and increases quadratically when the constraint is violated, thereby discouraging solutions that fall below the prescribed lower bound on $\bar{\sigma}_{\mathrm{yield}}$. In this work, $\alpha = 10$ is used.

For the target-matching inverse design case, the spall-strength and yield-strength matching errors are combined through a weighted scalarization:
\begin{equation}\label{eq:objective_C}
\min_{\mathbf{x} \in \Omega} \quad 
J(\mathbf{x}) = \left( \frac{\bar{\sigma}_{\mathrm{spall}}^{\mathrm{target}} - \bar{\sigma}_{\mathrm{spall}}(\mathbf{x})}{\bar{\sigma}_{\mathrm{spall}}^{\mathrm{target}}} \right)^2 
+ 
\alpha_y \left( \frac{\bar{\sigma}_{\mathrm{yield}}^{\mathrm{target}} - \bar{\sigma}_{\mathrm{yield}}(\mathbf{x})}{\bar{\sigma}_{\mathrm{yield}}^{\mathrm{target}}} \right)^2,
\end{equation}
where $\alpha_y > 0$ controls the relative importance of matching the yield strength. In this work, $\alpha_y = 1$ is used.

Starting from an initial simplex of $n+1$ vertices in an $n$-dimensional design space, the NM algorithm iteratively updates the simplex based on evaluations of $J(\mathbf{x})$. 
This derivative-free formulation is well suited to the present stochastic microstructural design problem. 
The initial design reported in the optimization results corresponds to the best-performing point within the initial simplex. Convergence is assessed based on changes in the objective function and simplex size. All candidate designs are maintained within the admissible design space $\Omega$. The results for the various design scenarios are presented in Section~\ref{sec:results}.

\section{Optimization results and discussion}\label{sec:results}
This section presents surrogate-driven optimization of polycrystalline microstructures for enhanced mechanical performance. The optimization studies are organized into three regimes representing different performance requirements. Section~\ref{sec:caseA} examines spall-strength maximization under a relaxed yield strength constraint. Section~\ref{sec:caseB} imposes a stricter yield strength requirement, thereby restricting the feasible design space. Section~\ref{sec:caseC} investigates target-matching optimization of both properties, enabling analysis of optimizer behavior under feasible and infeasible target conditions. Together, these cases provide a systematic view of microstructure design strategies under different performance requirements. The computational cost of the proposed framework is discussed in Section~\ref{sec:comp_cost}.

\subsection{Case A: $\bar{\sigma}_{\mathrm{spall}}$ maximization subject to $\bar{\sigma}_{\mathrm{yield}} \,\ge\, 40$ MPa}\label{sec:caseA}

This case examines optimization of microstructures to maximize $\bar{\sigma}_{\mathrm{spall}}$ under the constraint $\bar{\sigma}_{\mathrm{yield}} \ge 40$ MPa. The constrained problem is solved using the penalty-based objective defined in Eq.~\eqref{eq:max_J}.

The initial simplex is constructed from microstructures with small $d_{\mathrm{mean}}$, which correspond to high $\bar{\sigma}_{\mathrm{yield}}$. Initializing well above the prescribed yield strength threshold allows substantial evolution of both $\bar{\sigma}_{\mathrm{spall}}$ and $\bar{\sigma}_{\mathrm{yield}}$ during optimization. This setup reveals the intrinsic competition between resistance to plastic yielding and resistance to dynamic tensile failure, thereby clarifying the extent to which $\bar{\sigma}_{\mathrm{spall}}$ can be enhanced without violating the yield strength constraint.

\subsubsection{Full-space optimization in five dimensions}\label{sec:caseA1}

In Case~A1, the spall-strength maximization problem is solved in the full five-dimensional design space, with all microstructural design variables allowed to vary freely within their prescribed bounds summarized in Table~\ref{tab:design_bounds}. The nonlinear objective landscape defined in Eq.~\eqref{eq:max_J} is explored using the NM algorithm. The following analysis examines optimizer convergence through the evolution of $J$, changes in $\bar{\sigma}_{\mathrm{spall}}$ and $\bar{\sigma}_{\mathrm{yield}}$, and the evolution of $\overline{\mathrm{MMD}}$, followed by the optimization trajectories in the design space and associated microstructural changes. While $\overline{\mathrm{MMD}}$ monitors surrogate reliability during optimization, the mechanical performance of the resulting optimal design is independently verified using high-fidelity CPFE simulations.

\begin{figure*}[htbp]
\centering

\begin{minipage}{0.48\textwidth}
    \centering
    \textbf{Case A1}
\end{minipage}
\hfill
\begin{minipage}{0.48\textwidth}
    \centering
    \textbf{Case B}
\end{minipage}

\begin{subfigure}{0.48\textwidth}
    \includegraphics[width=\linewidth]{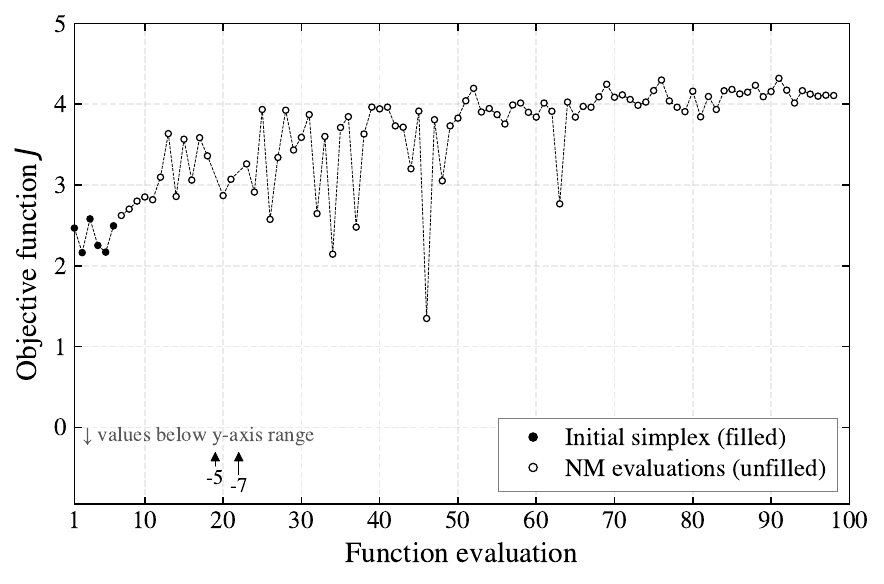}
    \caption{}
    \label{fig:J_caseA1}
\end{subfigure}
\hfill
\begin{subfigure}{0.48\textwidth}
    \includegraphics[width=\linewidth]{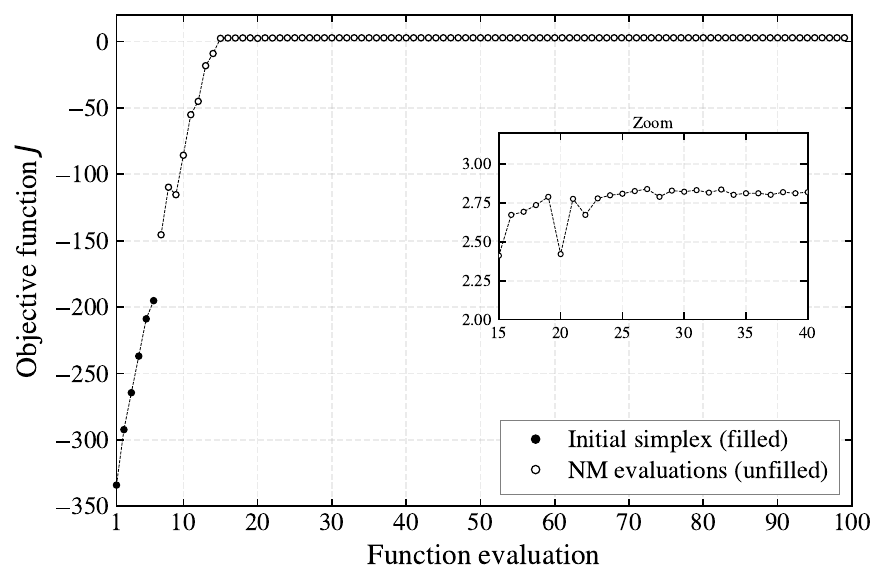}
    \caption{}
    \label{fig:J_caseB1}
\end{subfigure}

\begin{subfigure}{0.48\textwidth}
    \includegraphics[width=\linewidth]{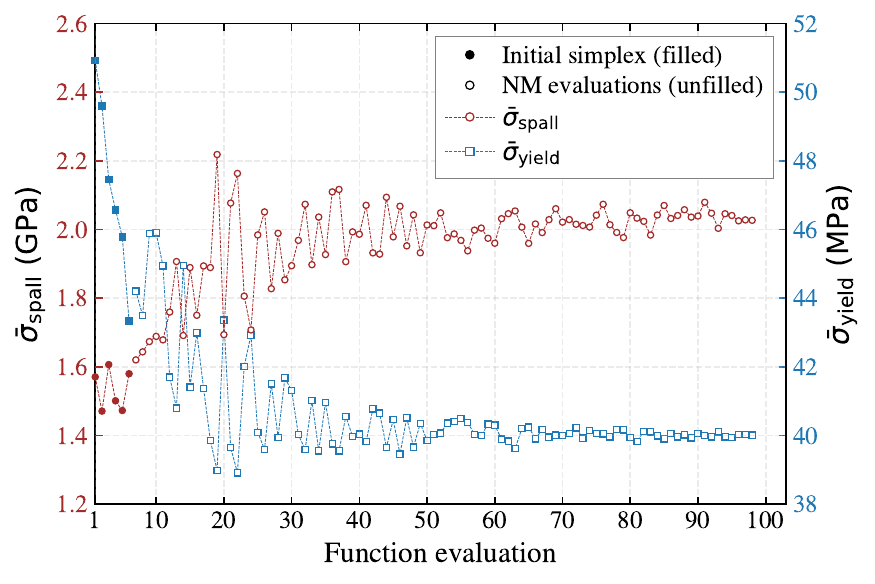}
    \caption{}
    \label{fig:spall_yield_twin_caseA1}
\end{subfigure}
\hfill
\begin{subfigure}{0.48\textwidth}
    \includegraphics[width=\linewidth]{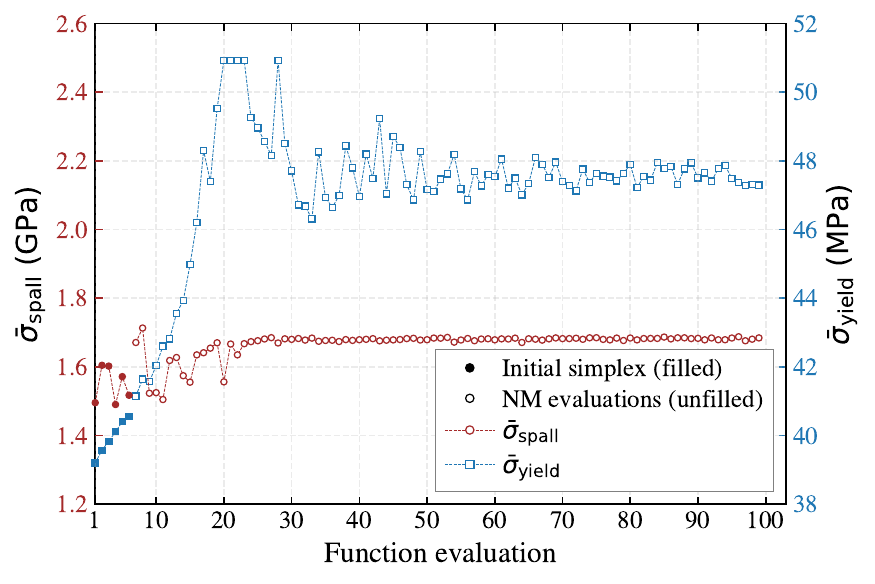}
    \caption{}
    \label{fig:spall_yield_twin_caseB1}
\end{subfigure}

\begin{subfigure}{0.48\textwidth}
    \includegraphics[width=\linewidth]{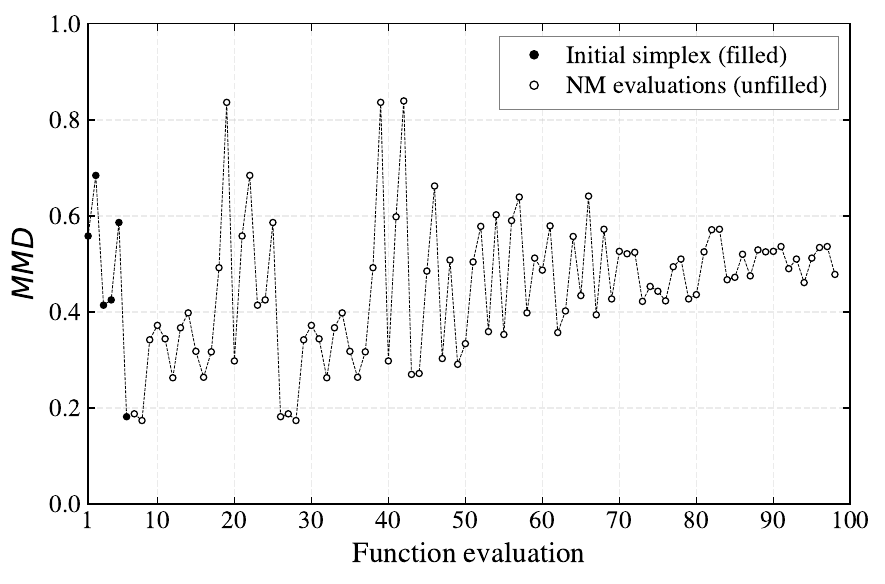}
    \caption{}
    \label{fig:MMD_caseA1}
\end{subfigure}
\hfill
\begin{subfigure}{0.48\textwidth}
    \includegraphics[width=\columnwidth]{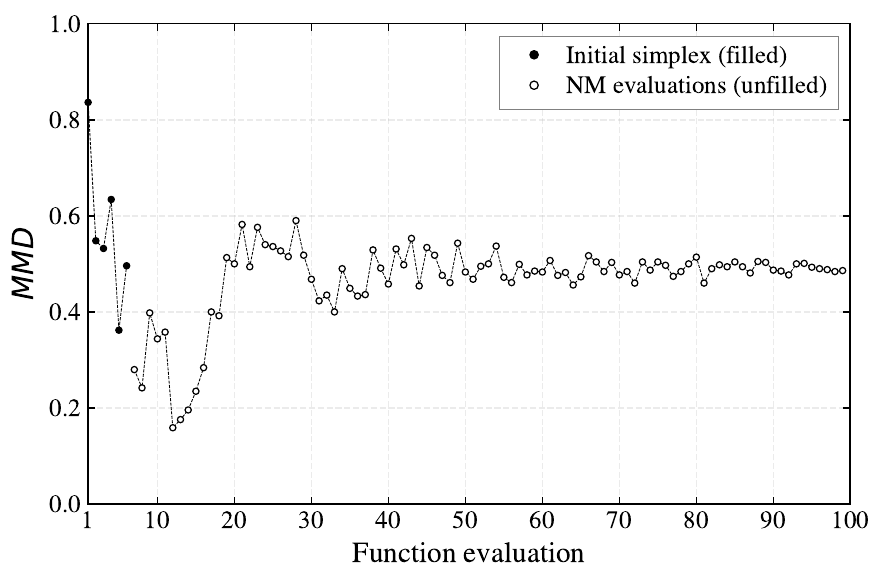}
    \caption{}
    \label{fig:MMD_caseB1}
\end{subfigure}

\caption{Evolution of optimization metrics over successive NM evaluations for Cases A1 and B. (a,b) Objective function \(J\); (c,d) \(\bar{\sigma}_{\mathrm{spall}}\) and \(\bar{\sigma}_{\mathrm{yield}}\); and (e,f) \(\overline{\mathrm{MMD}}\). Panels (a,c,e) correspond to Case A1 and panels (b,d,f) correspond to Case B. Filled markers denote the initial simplex points and unfilled markers denote subsequent evaluations. Values outside the plotted range in panel (a) are indicated by arrows and annotated numerically. The inset in panel (b) shows the late-stage evolution of \(J\) on an expanded scale.}
\label{fig:optimization_evolution}
\end{figure*}

Figure~\ref{fig:J_caseA1} shows the evolution of the objective function $J$. Beginning from the initial simplex, $J$ increases overall, reflecting progressive improvement in design performance.
Intermediate oscillations arise from exploratory simplex updates inherent to the NM algorithm.
Occasional violations of the yield strength constraint reduce $J$ through the quadratic penalty term. The objective subsequently stabilizes and approaches a clear plateau, indicating convergence to an optimal design within the prescribed bounds.

The evolution of $\bar{\sigma}_{\mathrm{spall}}$ and $\bar{\sigma}_{\mathrm{yield}}$ is shown in Figure~\ref{fig:spall_yield_twin_caseA1}. As $\bar{\sigma}_{\mathrm{spall}}$ increases, $\bar{\sigma}_{\mathrm{yield}}$ gradually decreases, demonstrating the inherent trade-off between spall and yield strengths under the imposed constraint. Rapid gains in $\bar{\sigma}_{\mathrm{spall}}$ occur during the early evaluations, whereas later evaluations progressively refine the balance between the two strengths as convergence is approached. Overall, $\bar{\sigma}_{\mathrm{spall}}$ increases by approximately $29\,\%$ relative to the initial design.

The distributional shift between the explored designs and the training distribution is monitored in Figure~\ref{fig:MMD_caseA1} using $\overline{\mathrm{MMD}}$. Although $\overline{\mathrm{MMD}}$ oscillates and temporarily exceeds the threshold $\tau = 0.55$ during intermediate exploratory steps, the final optimal design satisfies $\overline{\mathrm{MMD}} < \tau$. This confirms that the converged solution resides within the distribution-consistent reliability region $\Omega_\tau$ supported by the training data.

\begin{figure*}[t]
\centering
\begin{subfigure}{0.49\textwidth}
    \centering
    \includegraphics[width=\linewidth]{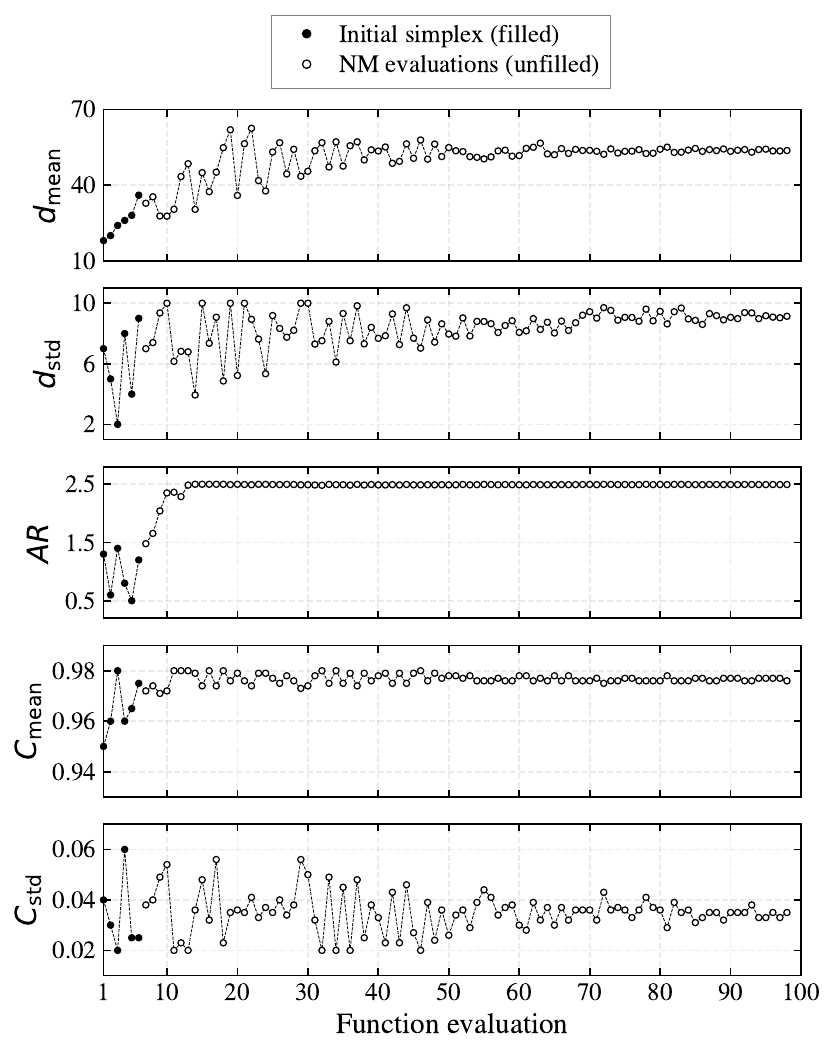}
    \caption{Case~A1}
    \label{fig:micro_descriptors_evolution_caseA1}
\end{subfigure}
\hfill
\begin{subfigure}{0.49\textwidth}
    \centering
    \includegraphics[width=\linewidth]{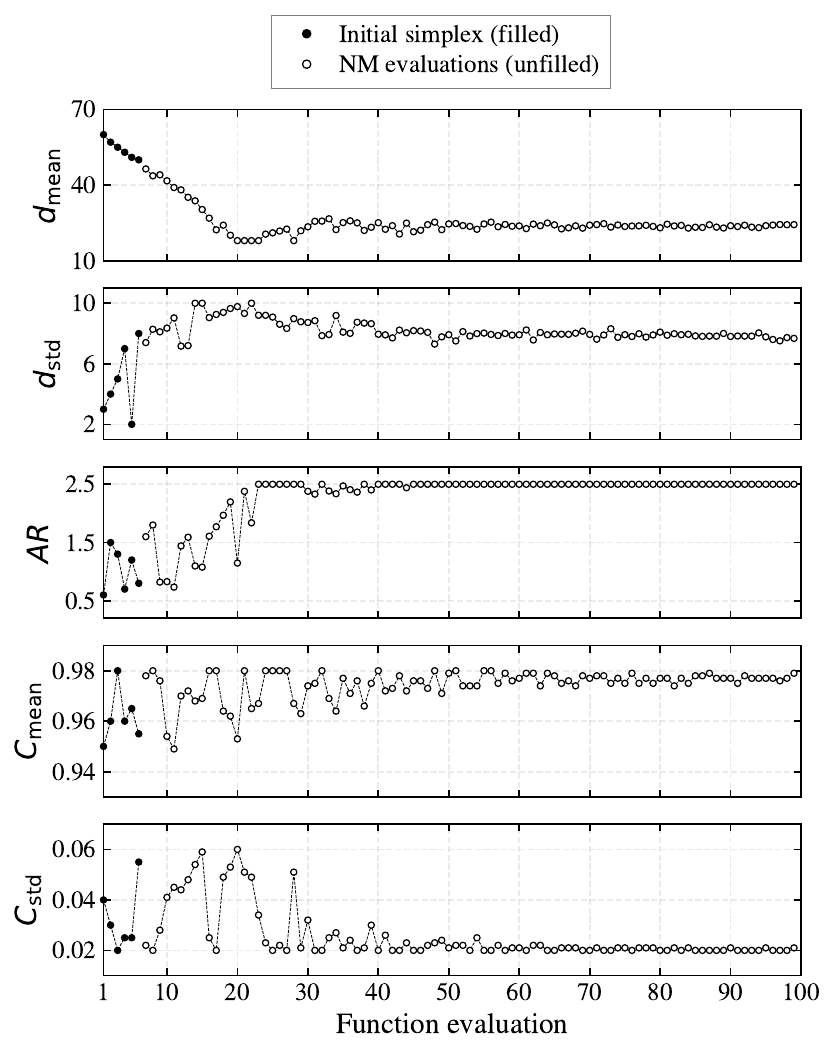}
    \caption{Case~B}
    \label{fig:micro_descriptors_evolution_caseB1}
\end{subfigure}
\caption{Evolution of the five microstructural design variables over successive NM evaluations for cases A1 and B. Filled markers denote the initial simplex points and unfilled markers denote subsequent evaluations.}
\label{fig:NM_evolution_combined_caseA}
\end{figure*}

Figure~\ref{fig:micro_descriptors_evolution_caseA1} illustrates the trajectories of the five design variables defined in Equation~\eqref{eq:design_vars}. The initially dispersed simplex progressively contracts toward a localized region of the design space characterized by large $d_{\mathrm{mean}}$, $d_{\mathrm{std}}$, and $AR$, corresponding to relatively large, elongated grains. All design variables subsequently stabilize to nearly constant values, consistent with the plateau observed in both the objective function and the mechanical responses.

\begin{figure*}[htbp]
\centering
\begin{subfigure}{0.7\columnwidth}
    \centering
    \includegraphics[width=\linewidth]{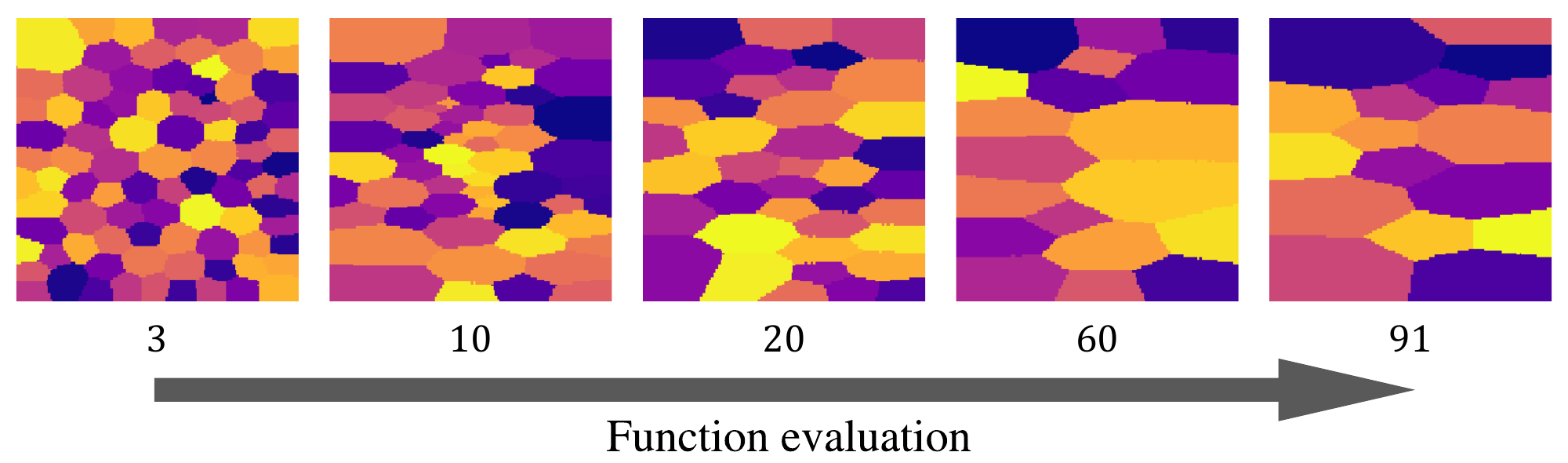}
    \caption{Case~A1}
    \label{fig:microstructure_evolution_caseA1}
\end{subfigure}
\par
\vspace{4mm}
\begin{subfigure}{0.7\columnwidth}
    \centering
    \includegraphics[width=\linewidth]{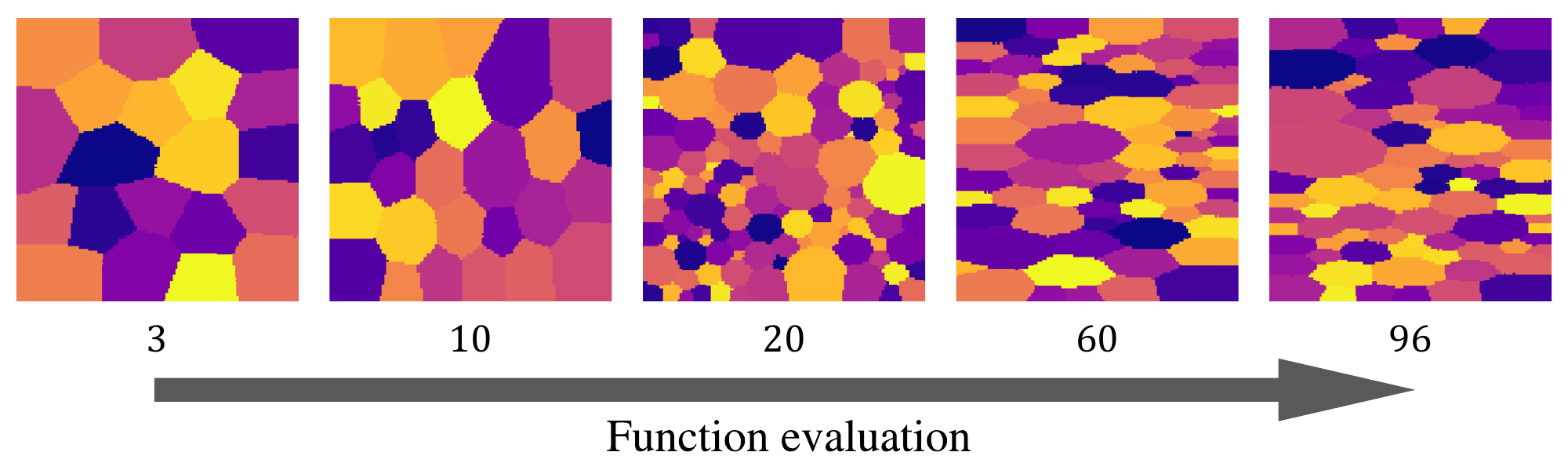}
    \caption{Case~B}
    \label{fig:microstructure_evolution_caseB1}
\end{subfigure}
\caption{Evolution of the microstructure during optimization over successive NM evaluations for Cases A1 and B. Representative samples at selected evaluations are shown, with evaluation numbers indicated below each image. For each case, the first image corresponds to the initial design and the last image shows the final optimal microstructure.}
\label{fig:microstructure_evolution_casesA_B}
\end{figure*}

\begin{figure*}[htbp]
\centering
\begin{subfigure}{0.7\columnwidth}
    \centering
    \includegraphics[width=\linewidth]{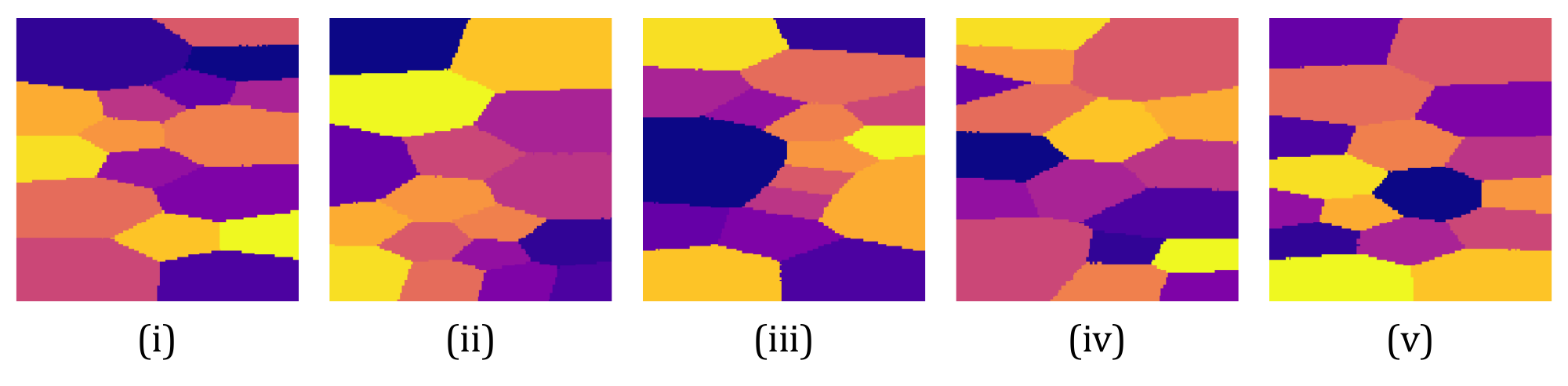}
    \caption{Case~A1}
    \label{fig:optimal_microstructures_caseA1}
\end{subfigure}
\par
\vspace{4mm}
\begin{subfigure}{0.7\columnwidth}
    \centering
    \includegraphics[width=\linewidth]{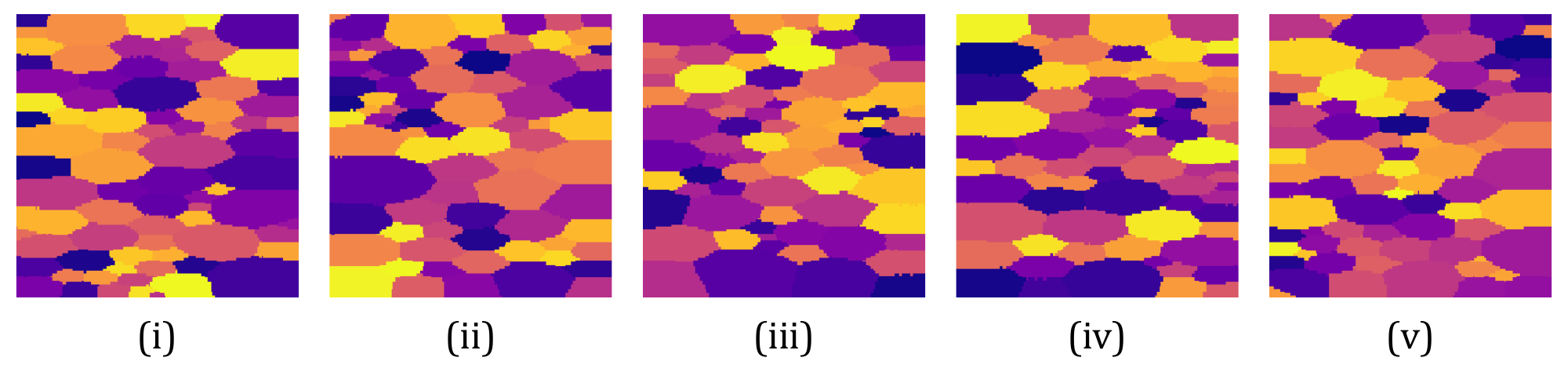}
    \caption{Case~B}
    \label{fig:optimal_microstructures_caseB1}
\end{subfigure}
\caption{Representative realizations of the optimal designs from ensembles of 100 statistically equivalent samples for Cases A1 and B. For each case, realizations (i)-(v) correspond to distinct random seeds $\xi_s$.}
\label{fig:optimal_microstructures_casesA1_B}
\end{figure*}

Figure~\ref{fig:microstructure_evolution_caseA1} presents representative microstructures sampled along the optimization trajectory, progressing from the initial design through intermediate designs to the final optimal design. Figure~\ref{fig:optimal_microstructures_caseA1} further illustrates that the optimal design represents an ensemble of stochastic microstructures, with distinct random seeds $\xi_s$ generating different realizations from the same optimal design vector.

\begin{figure*}[htbp]
\centering
\begin{subfigure}{0.7\columnwidth}
    \centering
    \includegraphics[width=\linewidth]{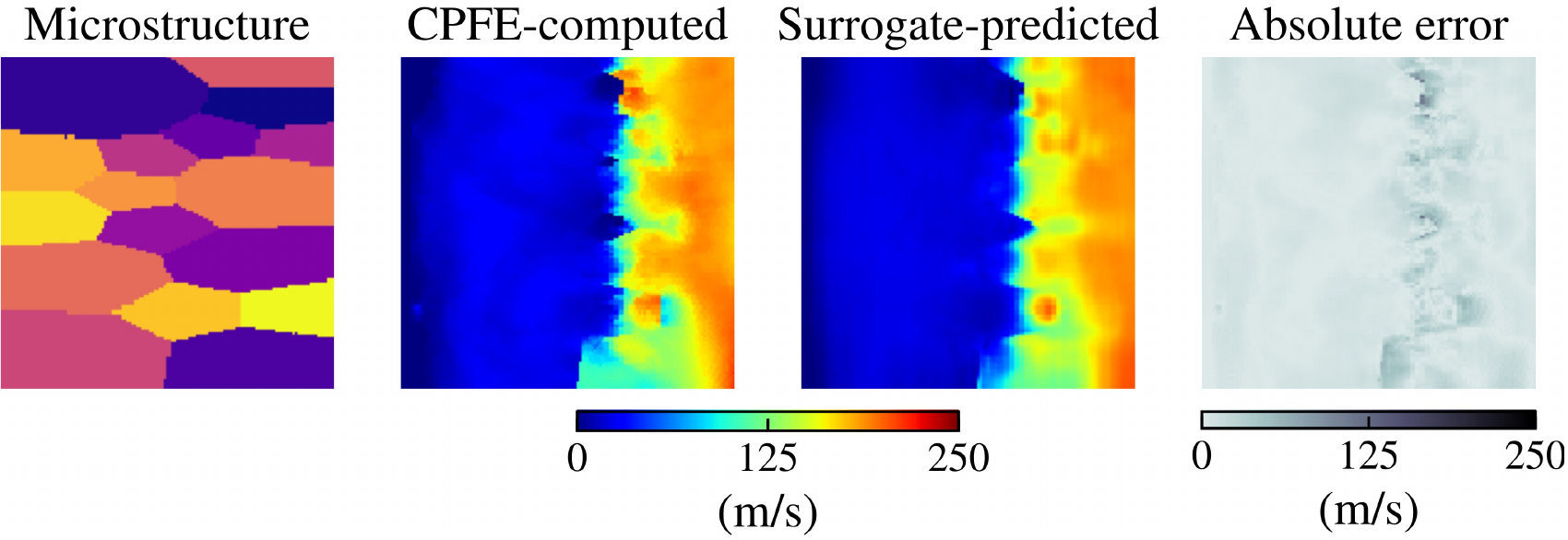}
    \caption{Case~A1}
    \label{fig:abaqus_error_caseA1}
\end{subfigure}
\par
\vspace{4mm}
\begin{subfigure}{0.7\columnwidth}
    \centering
    \includegraphics[width=\linewidth]{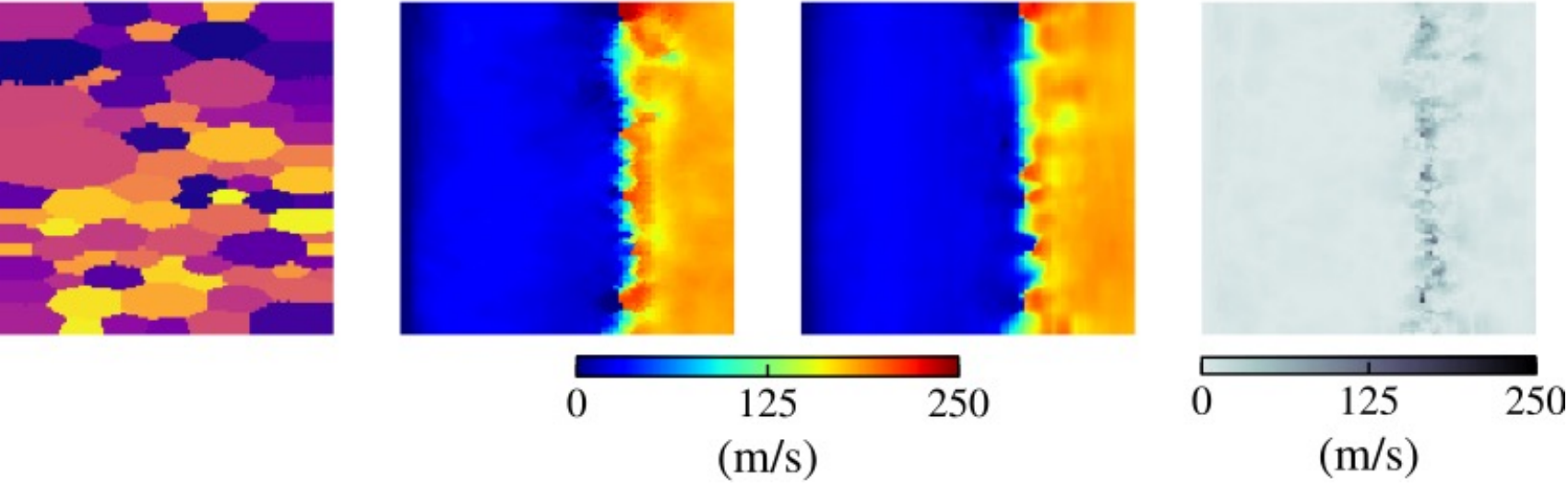}
    \caption{Case~B}
    \label{fig:abaqus_error_caseB1}
\end{subfigure}
\caption{CPFE-computed reference and surrogate-predicted velocity fields for representative optimal microstructures from Cases A1 and B at \(t=100\) ns following spall initiation, along with the corresponding absolute error fields.}\label{fig:abaqus_error_casesA_B}
\end{figure*}

The surrogate predictions for the optimal design are further validated against CPFE simulations. Figure~\ref{fig:abaqus_error_caseA1} compares the surrogate-predicted and CPFE-computed velocity fields at $t = 100$ ns following spall initiation, showing close overall agreement. Localized discrepancies are primarily confined to regions near grain boundaries along the spall plane away from the free surface, likely associated with the spatial averaging inherent in convolutional operations. Table~\ref{tab:true_pred_spall} shows that the relative error $\varepsilon_{\bar{\sigma}_{\mathrm{spall}}}$ in the ensemble-averaged $\bar{\sigma}_{\mathrm{spall}}$ over 100 realizations is approximately $3\,\%$, which is small compared to the $\sim 29\,\%$ improvement achieved through optimization. Notably, this accuracy is achieved at over two orders of magnitude lower computational cost (Section~\ref{sec:comp_cost}), enabling systematic large-scale design exploration that would otherwise be computationally prohibitive using direct CPFE simulations.

%

\begin{table}[htbp]
\centering
\caption{CPFE-computed reference and surrogate-predicted $\bar{\sigma}_{\mathrm{spall}}$ averaged over 100 stochastic realizations of the optimal microstructure, along with the corresponding relative error $\varepsilon_{\bar{\sigma}_{\mathrm{spall}}}\,(\%)$.}
\label{tab:true_pred_spall}
\begin{tabular}{c c c c}
\hline
\textbf{Case} &
\textbf{CPFE} $\bar{\sigma}_{\mathrm{spall}}$ (GPa) &
\textbf{Surrogate-predicted} $\bar{\sigma}_{\mathrm{spall}}$ (GPa) &
\textbf{$\varepsilon_{\bar{\sigma}_{\mathrm{spall}}}\,(\%)$} \\
\hline
\textbf{A1} & 2.01 & 2.07 & 3.0 \\
\textbf{B} & 1.67 & 1.69 & 1.2 \\
\hline
\end{tabular}
\end{table}

\begin{figure}[htbp]
\centering
\includegraphics[width=0.7\columnwidth,keepaspectratio]{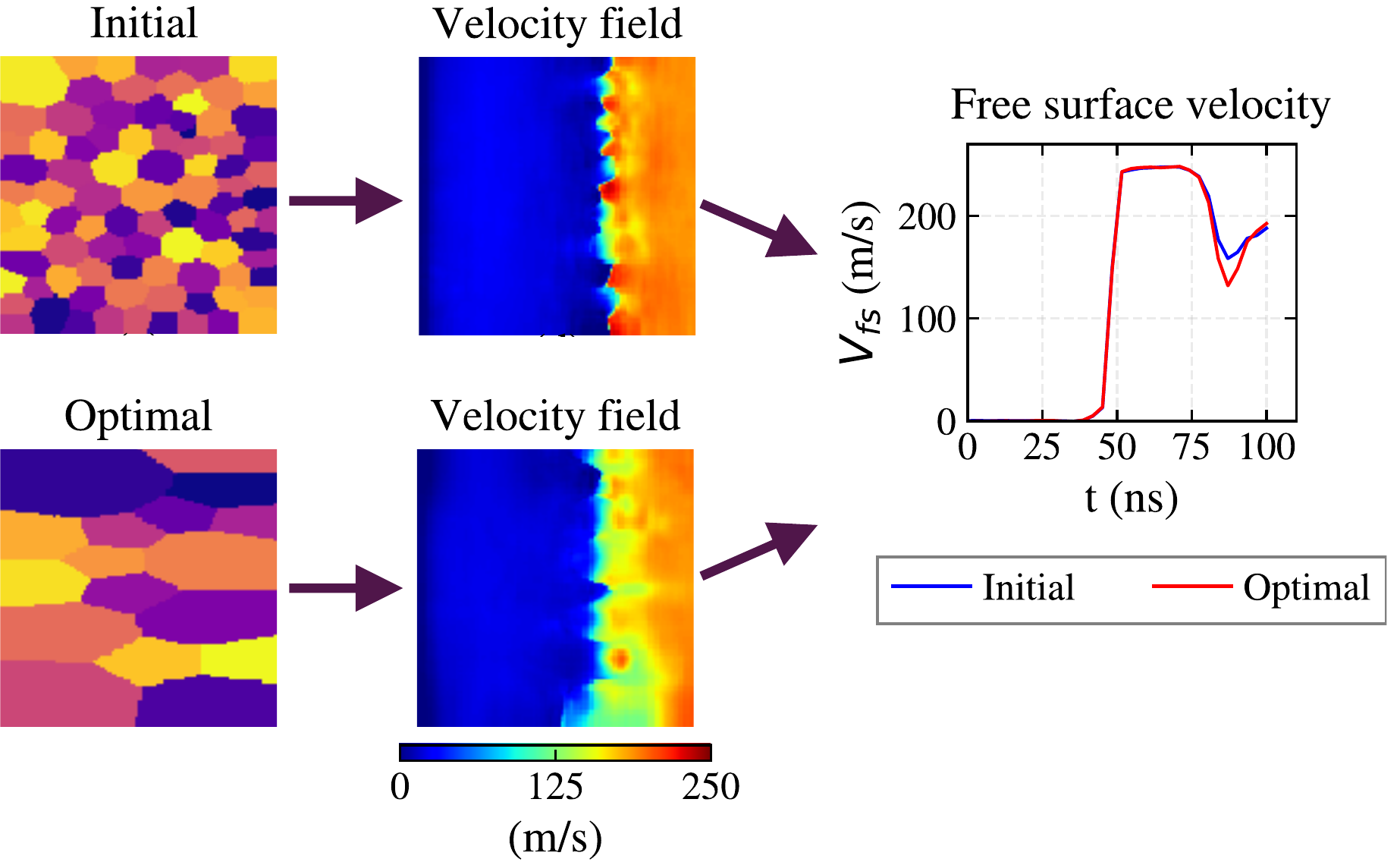}
\caption{Initial and optimal microstructures and their surrogate-predicted dynamic responses. Velocity fields are shown at \(t=100\) ns following spall initiation, together with the corresponding free-surface velocity histories.}
\label{fig:initial_vs_optimal_caseA1}
\end{figure}

To elucidate the mechanistic origin of the enhanced spall performance, the surrogate-predicted velocity field and free-surface velocity history $V_{fs}$ of the optimal microstructure are analyzed and compared with those of the initial design. Figure~\ref{fig:initial_vs_optimal_caseA1} presents the initial and optimal microstructures, their corresponding velocity fields at $t = 100$ ns following spall initiation, and the associated $V_{fs}$ histories. The optimal morphology modifies grain-boundary orientations relative to the loading direction, thereby governing interfacial separation and spall-plane formation. This results in a substantially stronger pull-back response compared to the initial design. The corresponding wave-interaction mechanism underlies the enhancement in $\bar{\sigma}_{\mathrm{spall}}$ achieved through optimization. 

The microstructure-resolved full-field predictions preserve spatial information essential for interpreting dynamic failure mechanisms. Unlike descriptor-to-strength regression models, which obscure intermediate physical processes, the present framework directly links microstructural morphology to tensile wave interactions, enabling mechanistic interpretation of spall behavior. Overall, the full-space optimization substantially enhances spall strength while satisfying the prescribed yield strength constraint, revealing the inherent trade-off between the two properties. Together, these results establish the proposed framework as an efficient approach for optimizing stochastic microstructures under competing mechanical requirements while retaining mechanistic insight into their performance.

\subsubsection{Subspace optimization with fixed variables}\label{sec:sub-caseA}

To further investigate the microstructure–property relationships identified in the full-space study, a series of reduced-dimensional (subspace) optimization problems are performed. In practice, processing routes may impose restrictions on microstructural design variables that limit independent control of certain descriptors. To account for such scenarios, selected design variables are fixed at prescribed values distinct from the full-space optimum, while the remaining variables vary freely within the same admissible bounds.

The constrained optimization problem defined in Eq.~\eqref{eq:max_J} is solved over the resulting subspace. This analysis serves two key purposes. First, it quantifies how fixing specific design variables limits the attainable $\bar{\sigma}_{\mathrm{spall}}$ and $\bar{\sigma}_{\mathrm{yield}}$ under the imposed yield strength constraint. Second, it clarifies whether the full-space optimum relies on strong coupling among all design variables or can be approximated by only a subset of dominant variables.

\begin{figure}[htbp]
\centering
\includegraphics[width=0.7\linewidth]{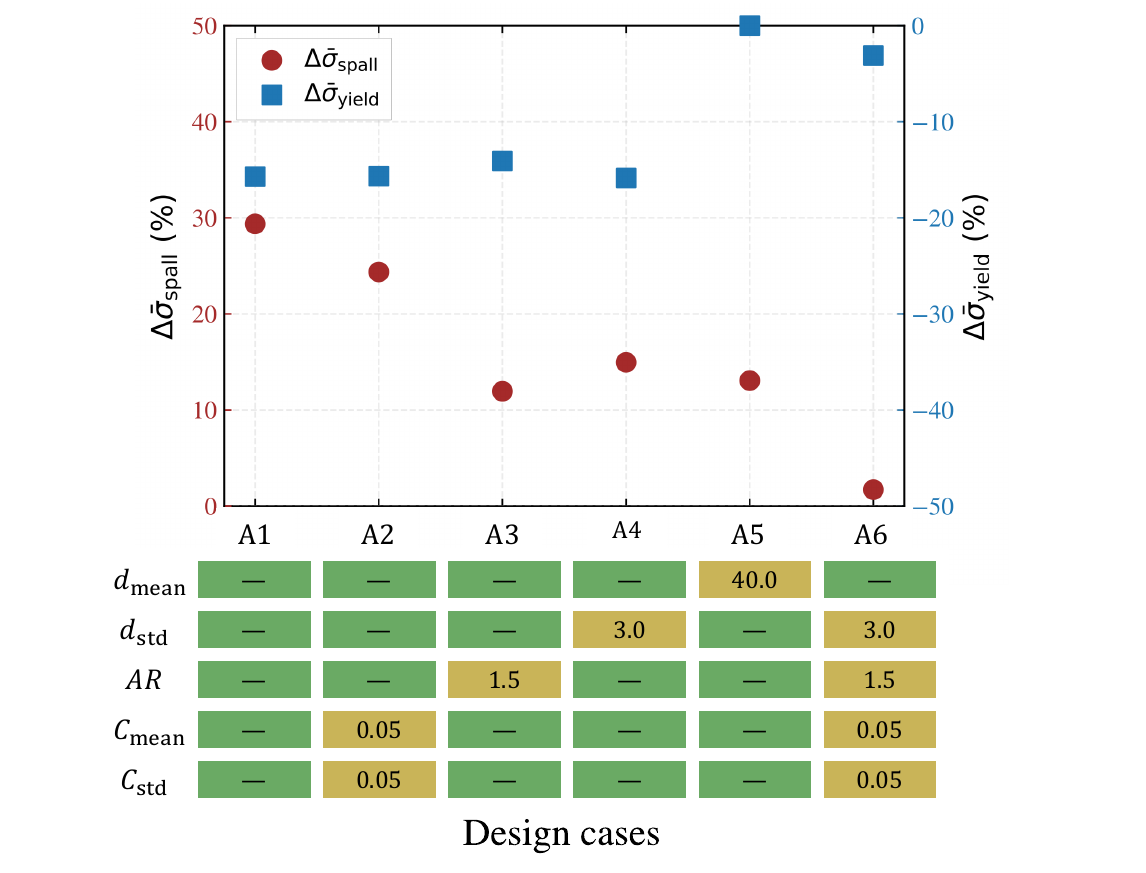}
\caption{Percentage change in spall and yield strengths, $\Delta \bar{\sigma}_{\mathrm{spall}}$ and $\Delta \bar{\sigma}_{\mathrm{yield}}$, relative to the initial design, for the six optimization subcases of Case~A. Fixed and free variables are listed for each subcase; "---" denotes a free variable and numerical entries indicate fixed values.}
\label{fig:scatter_optimal_design_cases_A}
\end{figure}

Six subcases are defined to isolate the influence of individual design variables. Case~A1 corresponds to the full five-dimensional optimization presented in Section~\ref{sec:caseA1}, where all design variables $(d_\mathrm{mean}, d_\mathrm{std}, AR, C_\mathrm{mean}, C_\mathrm{std})$ are free. In Case~A2, $C_\mathrm{mean}$ and $C_\mathrm{std}$ are fixed, restricting variations in grain circularity. Case~A3 fixes the aspect ratio $AR$, constraining grain elongation. Case~A4 fixes the grain size standard deviation $d_\mathrm{std}$, while Case~A5 fixes the mean grain size $d_\mathrm{mean}$. Case~A6 treats $d_\mathrm{mean}$ as the sole free variable, isolating its influence on mechanical performance.

Figure~\ref{fig:scatter_optimal_design_cases_A} shows the percentage changes in spall and yield strengths, $\Delta \bar{\sigma}_{\mathrm{spall}}$ and $\Delta \bar{\sigma}_{\mathrm{yield}}$, relative to the initial design for the six subcases. 
Case~A1 exhibits the largest increase in $\bar{\sigma}_{\mathrm{spall}}$, indicating that maximizing spall performance requires coordinated optimization of all design variables. This gain in $\bar{\sigma}_{\mathrm{spall}}$ is accompanied by a reduction in $\bar{\sigma}_{\mathrm{yield}}$, highlighting the strong trade-off between spall and yield strengths.
In contrast, Case~A6, where only $d_\mathrm{mean}$ is allowed to vary, shows negligible changes in both strengths, demonstrating that variation of $d_\mathrm{mean}$ alone is insufficient to drive significant performance improvement when the remaining variables are fixed. 
The intermediate subcases (Cases~A2–A5) exhibit smaller increases in $\bar{\sigma}_{\mathrm{spall}}$ compared with Case~A1 while still incurring comparable reductions in $\bar{\sigma}_{\mathrm{yield}}$. These results indicate that the full-space optimum cannot be approximated by optimizing only a subset of variables.
In Case~A5, where $d_\mathrm{mean}$ is fixed, no reduction in $\bar{\sigma}_{\mathrm{yield}}$ is observed since $\bar{\sigma}_{\mathrm{yield}}$ in the present model depends solely on $d_\mathrm{mean}$ through the Hall–Petch relation.

Overall, these results quantify the attainable spall enhancement under the imposed yield strength constraint and additional restrictions on selected microstructural design variables. The subspace analyses further demonstrate that the surrogate-driven framework enables efficient reduced-dimensional optimization and systematic large-scale design exploration that would otherwise be computationally prohibitive with direct CPFE simulations.

\subsection{Case B: $\bar{\sigma}_{\mathrm{spall}}$ maximization subject to $\bar{\sigma}_{\mathrm{yield}} \,\ge\, 45$ MPa}\label{sec:caseB}

Building on Case~A, which considered a relaxed yield strength requirement, Case~B investigates optimization under a stricter yield strength constraint. The objective is to maximize $\bar{\sigma}_{\mathrm{spall}}$ while satisfying $\bar{\sigma}_{\mathrm{yield}} \ge 45$ MPa. The constrained problem is solved using the penalty-based objective defined in Eq.~\eqref{eq:max_J}.

The initial simplex is constructed from microstructures with large $d_{\mathrm{mean}}$, corresponding to low $\bar{\sigma}_{\mathrm{yield}}$, such that the initial designs violate the prescribed yield strength constraint. Consequently, the optimizer must increase $\bar{\sigma}_{\mathrm{yield}}$ while simultaneously improving $\bar{\sigma}_{\mathrm{spall}}$, resulting in simultaneous strengthening of both properties. This initialization intensifies constraint activation during the early optimization steps and, together with the stricter yield strength requirement, provides a more restrictive optimization problem than Case~A. The $\bar{\sigma}_{\mathrm{spall}}$ maximization problem is solved in the same five-dimensional design space and with the same variable bounds as Case~A. The key differences arise from the stricter yield strength constraint and the infeasible initialization, which together narrow the feasible region and make constraint satisfaction a dominant driver of the optimization trajectory.

Figure~\ref{fig:J_caseB1} shows the evolution of the objective function $J$. Unlike Case~A1, where the initial simplex satisfies the yield strength constraint and $J$ increases monotonically aside from exploratory oscillations, the initial simplex in Case~B violates the stricter yield strength requirement, causing $J$ to initially take negative values due to the penalty term. As the constraint violation is progressively reduced, $J$ increases steadily and eventually approaches a plateau, indicating convergence to a locally optimal design within the feasible region.

Figure~\ref{fig:spall_yield_twin_caseB1} shows the evolution of $\bar{\sigma}_{\mathrm{spall}}$ and $\bar{\sigma}_{\mathrm{yield}}$. In contrast to Case~A1, where spall enhancement occurs alongside gradual yield reduction under the relaxed yield strength constraint, the stricter requirement in Case~B compels the optimizer to increase $\bar{\sigma}_{\mathrm{yield}}$ while maintaining high $\bar{\sigma}_{\mathrm{spall}}$, resulting in simultaneous strengthening of both properties. Rapid gains in yield strength dominate the early evaluations as as the optimizer moves toward the feasible region, followed by progressive refinement of both properties near convergence.

The distributional shift is monitored in Figure~\ref{fig:MMD_caseB1}. Several infeasible initial designs lie outside the distribution-consistent reliability region, with $\overline{\mathrm{MMD}}$ exceeding the threshold $\tau$. As optimization progresses, $\overline{\mathrm{MMD}}$ decreases, and both intermediate and final designs satisfy $\overline{\mathrm{MMD}} < \tau$. Thus, although the optimization begins with designs exhibiting substantial distributional shift, the converged solution lies within the distribution-consistent reliability region $\Omega_\tau$.

Figure~\ref{fig:micro_descriptors_evolution_caseB1} illustrates the trajectories of the five design variables. The initially dispersed simplex contracts toward a localized region characterized by smaller $d_{\mathrm{mean}}$ and larger $AR$, corresponding to thin, elongated grains. This convergence pattern differs from the large-grain-dominated region observed under the relaxed constraint in Case~A1 and reflects the shift toward grain refinement needed to increase $\bar{\sigma}_{\mathrm{yield}}$ and satisfy the stricter yield strength requirement. Figure~\ref{fig:microstructure_evolution_caseB1} presents representative microstructures along the optimization trajectory, illustrating the progressive evolution toward finer, elongated grains. Figure~\ref{fig:optimal_microstructures_caseB1} further illustrates different stochastic realizations of the optimal design generated using distinct random seeds $\xi_s$.

The surrogate predictions for the optimal design are further validated against CPFE simulations. Figure~\ref{fig:abaqus_error_caseB1} demonstrates close agreement between the surrogate-predicted and CPFE-computed velocity fields. Table~\ref{tab:true_pred_spall} shows that the relative error $\varepsilon_{\bar{\sigma}_{\mathrm{spall}}}$ in the ensemble-averaged $\bar{\sigma}_{\mathrm{spall}}$ over 100 realizations is approximately $1.2\,\%$, substantially smaller than the $\sim 11\,\%$ improvement achieved through optimization. The smaller increase in $\bar{\sigma}_{\mathrm{spall}}$ relative to the $29\,\%$ improvement obtained in Case~A1 reflects the more restrictive feasible region imposed by the stricter yield strength constraint.

\begin{figure}[htbp]
\centering
\includegraphics[width=0.7\columnwidth,keepaspectratio]{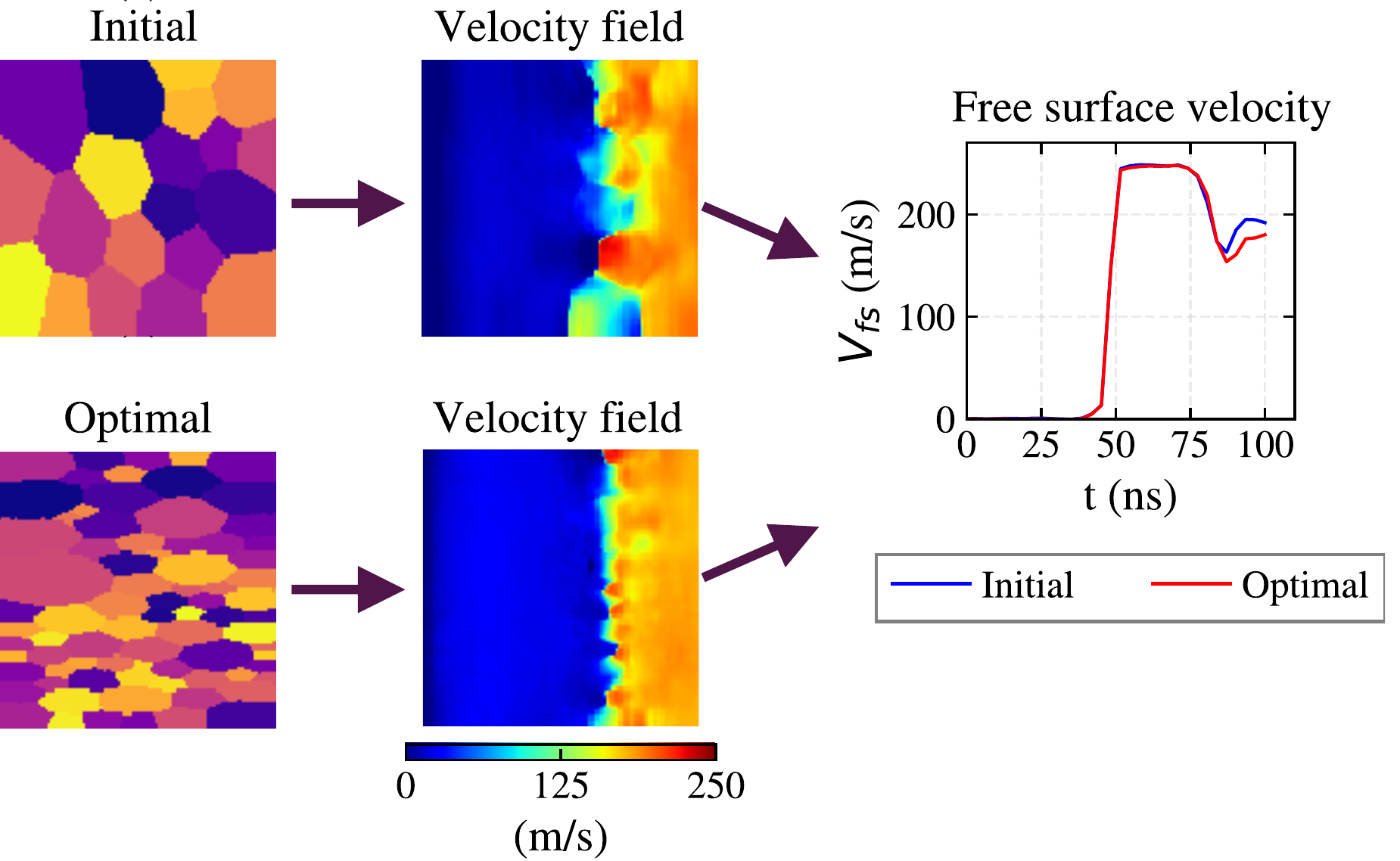}
\caption{Initial and optimal microstructures and their surrogate-predicted dynamic responses. Velocity fields are shown at \(t=100\) ns following spall initiation, together with the corresponding free-surface velocity histories.}
\label{fig:initial_vs_optimal_caseB1}
\end{figure}

To elucidate the mechanistic origin of the enhanced spall performance, the surrogate-predicted velocity field and free-surface velocity history ($V_{fs}$) of the optimal microstructure are analyzed and compared with those of the initial design. Figure~\ref{fig:initial_vs_optimal_caseB1} presents the corresponding microstructures, velocity fields at $t = 100$ ns following spall initiation, and the associated $V_{fs}$ histories. The optimal morphology, characterized by finer and elongated grains, alters grain-boundary orientations relative to the loading direction, influencing interfacial separation and spall-plane formation. The resulting changes in wave interactions produce a stronger pull-back response than the initial design, consistent with the enhanced $\bar{\sigma}_{\mathrm{spall}}$. However, the enhancement is more moderate than in Case~A1, reflecting the more restrictive design space imposed by the stricter yield strength constraint.

Overall, full-space optimization improves spall performance under the stricter yield strength constraint while requiring simultaneous strengthening of both mechanical properties. The converged design resides within the distribution-consistent reliability region, demonstrating the ability of the surrogate-driven framework to navigate competing mechanical requirements while maintaining reliable predictions.

\subsection{Case C: Target-matching inverse design of spall and yield strength}\label{sec:caseC}

This case investigates inverse design for simultaneously prescribed $\bar{\sigma}_{\mathrm{spall}}$ and $\bar{\sigma}_{\mathrm{yield}}$ targets. Unlike Cases~A and B, which maximize $\bar{\sigma}_{\mathrm{spall}}$ subject to a yield strength constraint, Case~C employs a target-matching objective that drives the design toward prescribed values of both properties. This formulation enables direct assessment of the optimizer's behavior when the prescribed targets are either attainable or unattainable within the design space, thereby distinguishing the convergence characteristics of feasible and infeasible target regimes.

\begin{figure*}[htbp]
\centering
\begin{subfigure}{0.49\columnwidth}
    \centering
    \includegraphics[width=\linewidth]{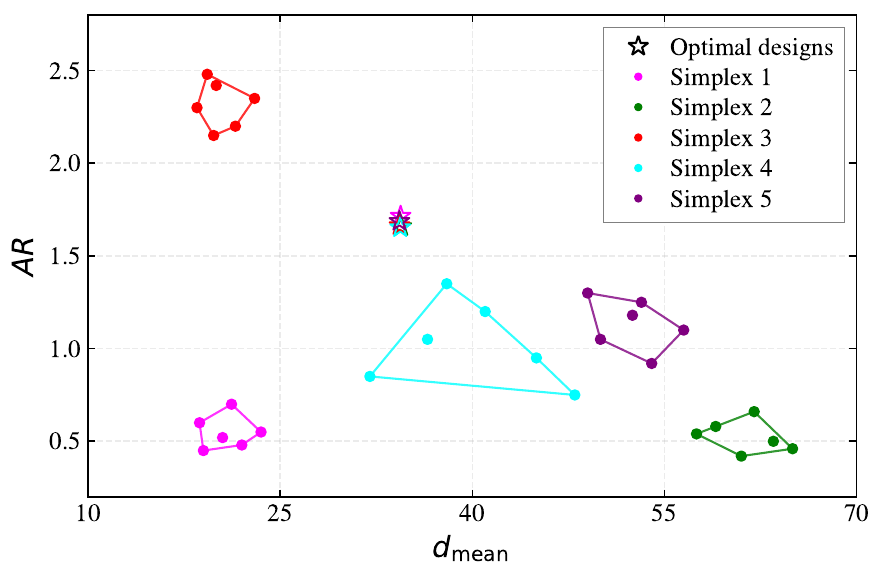}
    \caption{Case~C1}
    \label{fig:simplicesC1}
\end{subfigure}
\hfill
\begin{subfigure}{0.49\columnwidth}
    \centering
    \includegraphics[width=\linewidth]{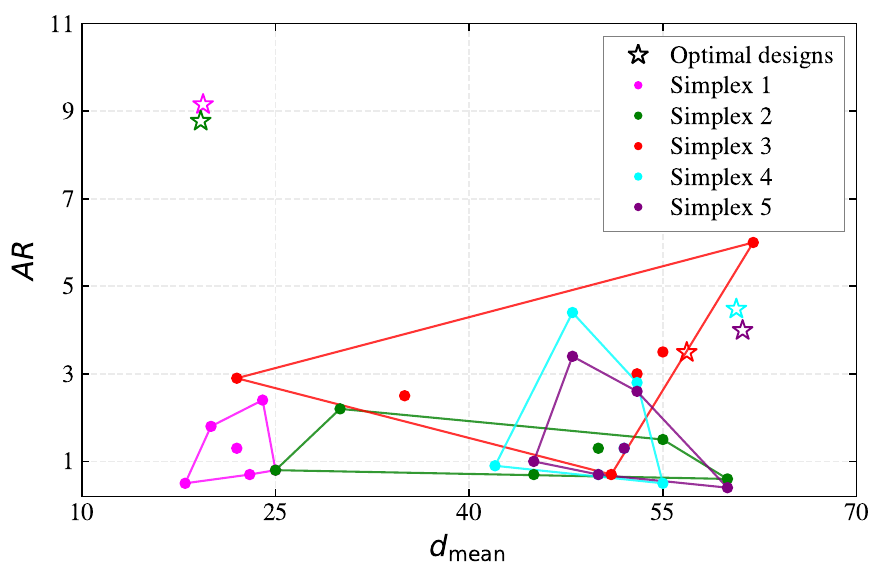}
    \caption{Case~C2}
    \label{fig:simplicesC2}
\end{subfigure}
\caption{Projection of the five-dimensional design space onto the ($d_{\mathrm{mean}}$, $AR$) subspace, showing five initial simplices and their corresponding converged designs. Filled markers denote the initial simplex vertices, with solid lines indicating the convex hull of each simplex. Starred markers denote the final optimal solutions obtained from the different initial simplices. (a) Case~C1: widely separated initial simplices converge to a tightly clustered solution, indicating a robust optimum largely insensitive to initialization. (b) Case~C2: an infeasible target produces distinct clusters of converged solutions, reflecting increased sensitivity to initialization.}
\label{fig:simplicesC1_C2}
\end{figure*}

\begin{figure*}[htbp]
\centering
\begin{subfigure}{0.7\columnwidth}
    \centering
    \includegraphics[width=\linewidth]{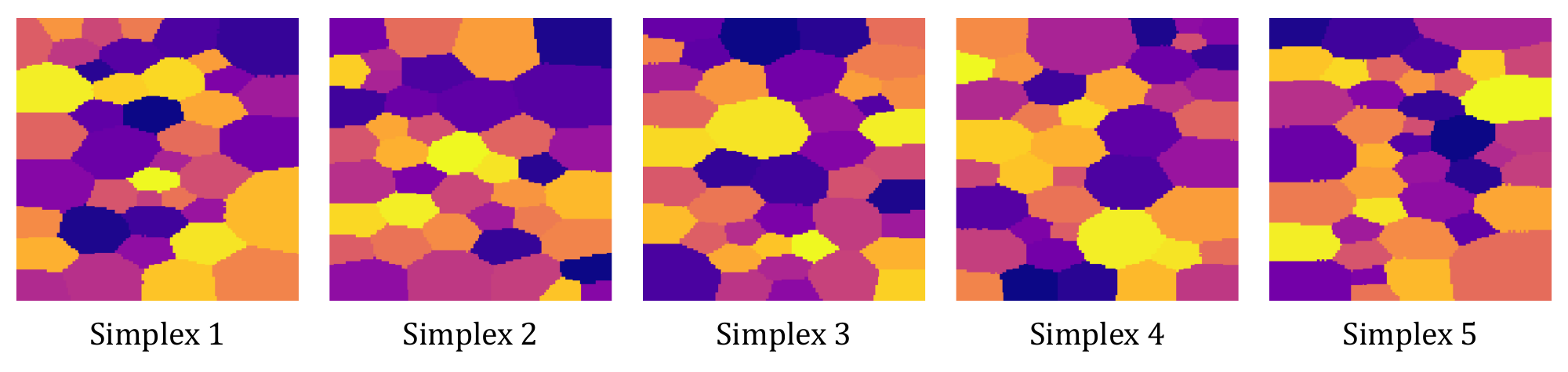}
    \caption{Case~C1}
    \label{fig:optimal_microstructures_subcaseC1}
\end{subfigure}
\par
\vspace{4mm}
\begin{subfigure}{0.7\columnwidth}
    \centering
    \includegraphics[width=\linewidth]{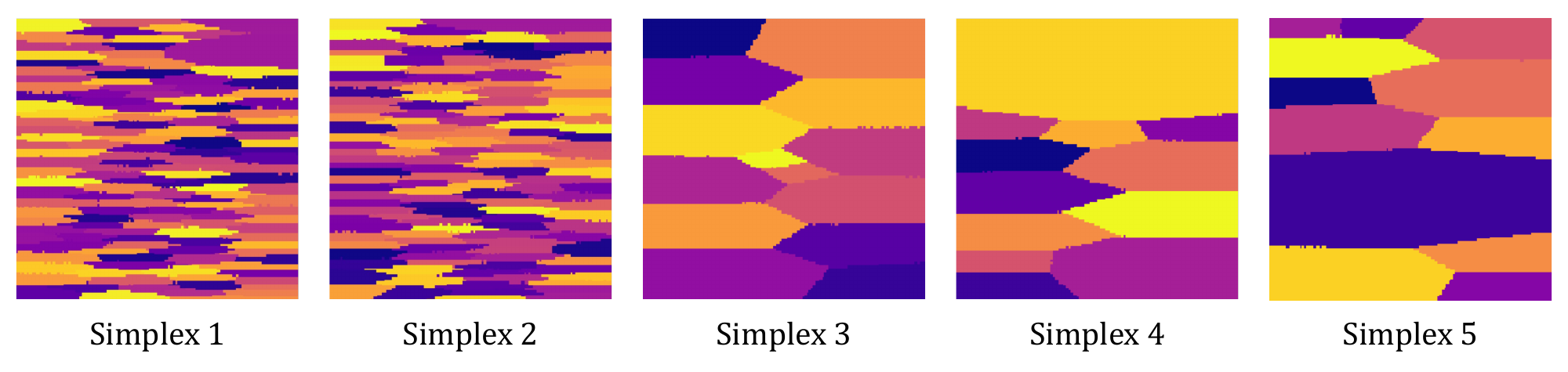}
    \caption{Case~C2}
    \label{fig:optimal_microstructures_subcaseC2}
\end{subfigure}
\caption{Representative optimal microstructures obtained from different initial simplices for Cases C1 and C2.}
\label{fig:optimal_microstructures_subcasesC1_C2}
\end{figure*}

\subsubsection{Case C1: Convergence to a feasible target}\label{sec:caseC1}

The target spall and yield strength combination is selected from a feasible region of the performance space identified in previous optimization runs. This subcase evaluates convergence behavior for a physically realizable and frequently attainable target. To assess robustness, the optimization is initialized from multiple distinct simplices. Figure~\ref{fig:simplicesC1} shows pairwise projections of five initial simplices and their corresponding optimal designs in the ($d_{mean}$, $AR$) subspace. Despite widely separated starting points, all optimization runs converge to a tightly clustered region in design space, indicating a robust optimum that is largely insensitive to the choice of initial simplex. Minor differences among the converged designs reflect local variability within a tightly confined region of the design space, consistent with convergence to a well-defined optimum with effectively equivalent performance. Figure~\ref{fig:optimal_microstructures_subcaseC1} presents representative optimal microstructures obtained from different initial simplices, demonstrating convergence toward similar morphologies. Low $\overline{\mathrm{MMD}}$ values for the optimal microstructures, reported in Table~\ref{tab:mmd_simplices}, confirm that the converged solutions lie within $\Omega_\tau$.

\begin{table}[htbp]
\centering
\caption{$\overline{\mathrm{MMD}}$ values for optimal microstructures obtained from different initial simplices for Cases C1 and C2.}
\label{tab:mmd_simplices}
\begin{tabular}{c c c c c c}
\hline
 \textbf{Case} & \textbf{Simplex 1} & \textbf{Simplex 2} & \textbf{Simplex 3} & \textbf{Simplex 4} & \textbf{Simplex 5} \\
\hline
\textbf{C1} & 0.185 & 0.184 & 0.186 & 0.180 & 0.186 \\
\textbf{C2} & 0.844 & 0.844 & 0.762 & 0.920 & 0.960 \\
\hline
\end{tabular}
\end{table}

\subsubsection{Case C2: Convergence behavior with an infeasible target}\label{sec:caseC2}

The prescribed target combination of spall and yield strength is intentionally selected from an infeasible region of the performance space identified in previous optimization runs. As a result, the optimizer cannot converge to the specified target and instead seeks the design that minimizes the target-matching objective defined in Eq.~\ref{eq:objective_C}. To examine optimizer behavior beyond the design space considered in the preceding cases, the bounds on the microstructural design variables are significantly expanded relative to the training domain, as summarized in Table~\ref{tab:design_bounds_C2}. These extended bounds admit microstructures that deviate substantially from those represented in the training distribution. This departure is reflected in the large $\overline{\mathrm{MMD}}$ values observed for the resulting optimal designs, as reported in Table~\ref{tab:mmd_simplices}.

\begin{table}[htbp]
\centering
\caption{Bounds on microstructural design variables for the Case~C2.}
\label{tab:design_bounds_C2}
\begin{tabular}{l
                S[table-format=2.3]
                S[table-format=2.3]}
\hline
Variable & {Lower bound} & {Upper bound} \\
\hline
$d_{\mathrm{mean}}$ ($\mu\mathrm{m}$) & 10.00 & 100.00 \\
$d_{\mathrm{std}}$ ($\mu\mathrm{m}$)  &  1.0 & 20.00 \\
$AR$                                  &  0.10 &  10.00 \\
$C_{\mathrm{mean}}$                   &  0.940 &  0.980 \\
$C_{\mathrm{std}}$                    &  0.020 &  0.060 \\
\hline
\end{tabular}
\end{table}

Figure~\ref{fig:simplicesC2} shows pairwise projections of five initial simplices and their corresponding optimal designs in the ($d_{\mathrm{mean}}$, $AR$) subspace. In contrast to the feasible-target case, the optimization trajectories terminate in distinct regions of the design space, indicating that the converged solution depends on the choice of initial simplex. The resulting variability among the converged designs reveals multiple locally optimal microstructural solutions that provide different compromises in approaching the unattainable target, highlighting the non-uniqueness of the inverse microstructure-property mapping under infeasible performance requirements. Figure~\ref{fig:optimal_microstructures_subcaseC2} presents representative optimal microstructures obtained from different initial simplices, further illustrating convergence toward distinct morphological regions.

\begin{figure}[htbp]
\centering
\includegraphics[width=0.5\columnwidth]{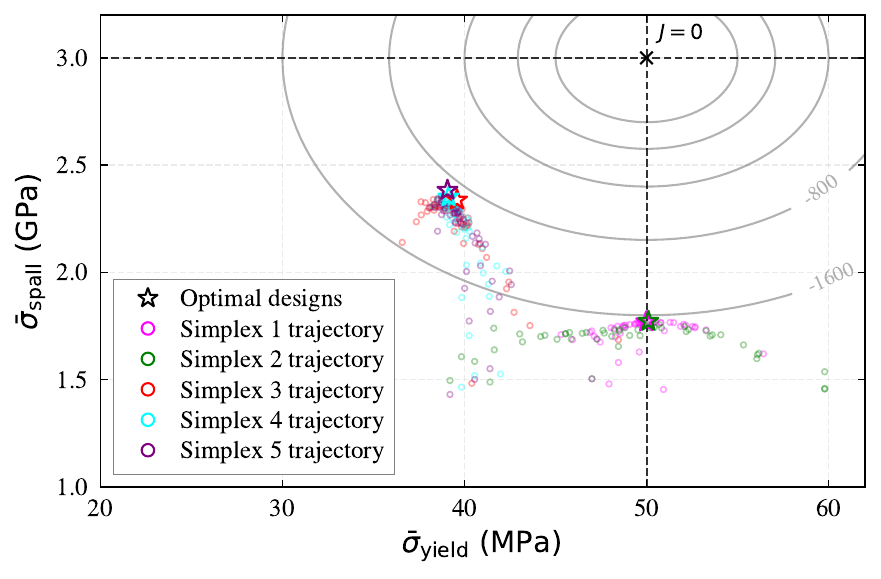}
\caption{$\bar{\sigma}_{\mathrm{spall}}$-$\bar{\sigma}_{\mathrm{yield}}$ performance map for Case~C2 under the infeasible target. Colored point clouds show the optimization trajectories from different initial simplices, while star markers denote the corresponding converged designs. Dashed lines indicate the prescribed targets for $\bar{\sigma}_{\mathrm{spall}}$ and $\bar{\sigma}_{\mathrm{yield}}$, and gray contours represent iso-levels of the quadratic target-matching objective $J$. The converged designs form distinct clusters corresponding to local optima, highlighting the trade-off between spall and yield performance.}
\label{fig:C2_strength_map}
\end{figure}

Figure~\ref{fig:C2_strength_map} shows the$\bar{\sigma}_{\mathrm{yield}}$-$\bar{\sigma}_{\mathrm{spall}}$ performance map for the converged designs obtained from different initial simplices under the infeasible Case~C2 target. In contrast to the feasible Case~C1 setting, the solutions do not converge to a single region but instead partition into two distinct clusters. These clusters correspond to designs that favor either higher $\bar{\sigma}_{\mathrm{spall}}$ at reduced $\bar{\sigma}_{\mathrm{yield}}$ or higher $\bar{\sigma}_{\mathrm{yield}}$ at reduced $\bar{\sigma}_{\mathrm{spall}}$. The absence of solutions in the upper-right quadrant defined by the prescribed thresholds is consistent with the joint target being unattainable within the considered design space. This behavior highlights the inherent trade-off between spall and yield performance and demonstrates increased sensitivity to initialization when targeting an infeasible combination of properties.

Taken together, the three optimization studies reveal how imposed property constraints alter the microstructural design landscape. Relaxed constraints permit spall enhancement through grain morphology changes at the expense of yield strength, whereas stricter constraints require coordinated variation of multiple descriptors to achieve concurrent strengthening. When both properties are prescribed simultaneously, feasible targets lead to consistent convergence, whereas infeasible targets produce distinct locally optimal designs representing different compromises between spall and yield strength. These results demonstrate the ability of the proposed framework to navigate stochastic microstructural design spaces under competing strength requirements while revealing the trade-offs and feasibility limits that govern attainable performance.

\subsection{Computational efficiency}\label{sec:comp_cost}
The computational cost of the proposed deep learning-driven optimization framework is compared with an equivalent high-fidelity CPFE-based implementation in terms of evaluation time. Each optimization run consists of approximately 100 objective function $J$ evaluations, with 100 stochastic microstructure realizations per $J$ evaluation, resulting in a total of $\sim 10{,}000$ microstructure evaluations per run. Microstructure generation using \textit{Neper} requires $\sim 5$ minutes per 100 microstructures. Because microstructure generation is common to both approaches, the following comparison focuses on CPFE evaluation and surrogate inference times. Each CPFE simulation is performed using 32 CPUs and requires $\sim 30$ minutes, such that the $\sim 10{,}000$ microstructure evaluations required for a single CPFE-driven optimization run correspond to $\sim 5{,}000$ hours of cumulative simulation time ($\approx 7$ months). In contrast, the surrogate model evaluates a batch of 100 microstructures in $\sim 30$ seconds on an NVIDIA A100 GPU, corresponding to a cumulative inference time of $\sim 0.83$ hours ($\approx 50$ minutes) for the same $\sim 10{,}000$ evaluations. Thus, replacing direct CPFE simulations with surrogate inference reduces the cumulative evaluation time by a factor of approximately $6{,}000$. Although concurrent CPFE simulations could reduce the elapsed wall-clock time, large-scale parallelization in Abaqus is additionally constrained by licensing limits on the number of simultaneous solver instances.

The surrogate workflow incurs a one-time development cost of $\sim 30$ minutes for the generation of 600 microstructures, $\sim 300$ hours for the corresponding CPFE simulations, and $\sim 8$ hours for model training. The break-even point, defined as the number of evaluations beyond which surrogate development and inference become computationally less expensive than direct CPFE evaluation, is estimated following the analysis in~\cite{di2023neural}. The resulting break-even point is approximately 617 evaluations, substantially below the $\sim 10{,}000$ evaluations required for a single optimization run.

The evaluation time per optimization run is summarized in Table~\ref{tab:cost_comparison}. Across the 17 optimization runs performed in this study, an equivalent CPFE-driven analysis would require approximately $85{,}000$ hours of cumulative simulation time. Conducting the full optimization study using direct CPFE simulations would therefore be computationally prohibitive. By replacing repeated CPFE simulations with surrogate inference, the proposed framework reduces the cumulative evaluation time by approximately four orders of magnitude, enabling large-scale, systematic optimization of the microstructural design space.

\begin{table}[htbp]
\centering
\caption{Approximate evaluation time per optimization run. Surrogate inference is performed on an NVIDIA A100 GPU, whereas each CPFE simulation uses 32 CPUs.}
\begin{tabular}{lccc}
\hline
Approach & Evaluations/run & Time/evaluation & Total time \\
\hline
Surrogate & $\sim 10{,}000$ & $\sim 0.3$ s & $\sim 0.83$ h \\
Equivalent CPFE & $\sim 10{,}000$ & $\sim 30$ min & $\sim 5{,}000$ h \\
\hline
\end{tabular}
\label{tab:cost_comparison}
\end{table}

\section{Conclusions}\label{sec:conclusions}
This work presents a deep learning-driven optimization framework for the design of polycrystalline microstructures under competing quasi-static and dynamic performance requirements. The framework couples stochastic microstructure generation with a full-field surrogate model and derivative-free optimization, enabling efficient design of stochastic grain morphologies that would be prohibitively expensive using direct CPFE-based optimization. A distribution-consistent reliability region (DCRR), based on maximum mean discrepancy (MMD), is used to monitor distributional shift between candidate and training microstructures as the optimizer navigates the design space.

The optimization studies reveal distinct design regimes associated with the competing requirements of spall and yield strength. 
Under relaxed yield strength constraints, spall enhancement occurs primarily through morphological changes that increase spall resistance at the expense of yield strength. 
Under stricter constraints, concurrent strengthening of both properties becomes necessary, restricting the feasible design space and increasing the influence of grain morphology, particularly aspect ratio. 
Subspace analyses further demonstrate that coordinated variation of multiple microstructural descriptors is required to achieve substantial improvements in performance. The optimal microstructures achieve up to approximately 29\% enhancement in mean spall strength while satisfying the prescribed yield strength constraint. Target-matching inverse design further demonstrates the ability to recover microstructural designs for prescribed combinations of spall and yield strength.

The surrogate accurately predicts full-field velocity histories and corresponding spall strengths across a broad range of microstructural descriptors, with the optimized designs validated against independent CPFE simulations with errors below 3$\%$. 
Beyond enabling rapid evaluation, the full-field predictions retain spatially and temporally resolved information linking grain morphology to wave interactions and dynamic failure, providing physical insight into why particular microstructures improve spall performance. 
Replacing repeated high-fidelity simulations with surrogate evaluations reduces the cumulative evaluation time by approximately four orders of magnitude.

Overall, the results demonstrate that surrogate-based optimization can efficiently navigate stochastic microstructural design spaces while accounting for competing mechanical performance requirements and monitoring surrogate reliability during optimization. 
Although demonstrated here for polycrystalline copper under plate-impact loading, the broader framework is applicable to optimization problems in which high-fidelity full-field simulations are prohibitively expensive but spatially and temporally resolved response information is essential both for evaluating performance and for understanding the physical mechanisms underlying optimal designs. 
Future extensions could integrate high-throughput dynamic experiments with the proposed computational framework, enabling predictive models to guide experiments toward promising regions of the materials design space and ultimately accelerate microstructure design~\cite{pyzer2022accelerating,hufnagel2026aimd}.

\section*{Acknowledgments}
Research was accomplished through the "AI-Driven Integrated and Automated Materials Design for Extreme Environments (AMDEE)" project sponsored by the Army Research Laboratory under Cooperative Agreement Number W911NF-23-2-0062. The views and conclusions contained in this document are those of the authors and should not be interpreted as representing the official policies, either expressed or implied, of the Army Research Laboratory or the U.S. Government. The U.S. Government is authorized to reproduce and distribute reprints for Government purposes notwithstanding any copyright notation herein. The authors gratefully acknowledge Dr. Piyush Wanchoo, and Dr. Rohit Berlia at Hopkins Extreme Materials Institute (HEMI) for the many discussions and invaluable insights that shaped these results.

This work was carried out at the Advanced Research Computing at Hopkins (ARCH) core facility (rockfish.jhu.edu), which is supported by the National Science Foundation (NSF) grant number OAC1920103.

\appendix
 
\bibliography{reference}

@book{antoun2003spall,
	author = {Antoun, Tarabay and Curran, Donald R and Razorenov, Sergey V and Seaman, Lynn and Kanel, Gennady I and Utkin, Alexander V},
	publisher = {Springer},
	title = {Spall fracture},
	year = {2003}}

@article{mcdowell2008concurrent,
	author = {McDowell, David L and Olson, Gregory B},
	journal = {Scientific Modeling and Simulation SMNS},
	number = {1},
	pages = {207--240},
	publisher = {Springer},
	title = {Concurrent design of hierarchical materials and structures},
	volume = {15},
	year = {2008}}

@article{mallick2021spall,
	author = {Mallick, Debjoy D and Prameela, Suhas Eswarappa and Ozturk, Deniz and Williams, Cyril L and Kang, Minju and Valentino, Gianna M and Lloyd, Jeffrey T and Wilkerson, Justin W and Weihs, Timothy P and Ramesh, KT},
	journal = {Mechanics of Materials},
	pages = {104065},
	publisher = {Elsevier},
	title = {Spall strength in alloyed magnesium: A compendium of research efforts from the CMEDE 10-year effort},
	volume = {162},
	year = {2021}}

@article{ragan2026microstructural,
	author = {Ragan, Tyler and Hao, Tengyuan and Zhou, Min},
	journal = {International Journal of Fracture},
	number = {2},
	pages = {16},
	publisher = {Springer},
	title = {Microstructural mechanisms for enhancing energy dissipation and spall resistance in SiC: T. Ragan et al.},
	volume = {250},
	year = {2026}}

@article{minich2004effect,
	author = {Minich, Roger W and Cazamias, James U and Kumar, Mukui and Schwartz, Adam J},
	journal = {Metallurgical and Materials Transactions A},
	number = {9},
	pages = {2663--2673},
	publisher = {Springer},
	title = {Effect of microstructural length scales on spall behavior of copper},
	volume = {35},
	year = {2004}}

@article{wilkerson2016unraveling,
	author = {Wilkerson, JW and Ramesh, KT},
	journal = {Physical review letters},
	number = {21},
	pages = {215503},
	publisher = {APS},
	title = {Unraveling the anomalous grain size dependence of cavitation},
	volume = {117},
	year = {2016}}

@article{escobedo2011effects,
	author = {Escobedo, JP and Dennis-Koller, Darcie and Cerreta, EK and Patterson, BM and Bronkhorst, CA and Hansen, BL and Tonks, D and Lebensohn, RA},
	journal = {Journal of Applied Physics},
	number = {3},
	publisher = {AIP Publishing},
	title = {Effects of grain size and boundary structure on the dynamic tensile response of copper},
	volume = {110},
	year = {2011}}

@article{fromm2009grain,
	author = {Fromm, Bradley S and Adams, Brent L and Ahmadi, Sadegh and Knezevic, Marko},
	journal = {Acta Materialia},
	number = {8},
	pages = {2339--2348},
	publisher = {Elsevier},
	title = {Grain size and orientation distributions: Application to yielding of $\alpha$-titanium},
	volume = {57},
	year = {2009}}

@article{abdolvand2018strong,
	author = {Abdolvand, Hamidreza and Wright, Jonathan and Wilkinson, Angus J},
	journal = {Nature communications},
	number = {1},
	pages = {171},
	publisher = {Nature Publishing Group UK London},
	title = {Strong grain neighbour effects in polycrystals},
	volume = {9},
	year = {2018}}

@article{haouala2018analysis,
	author = {Haouala, Sarra and Segurado, Javier and LLorca, Javier},
	journal = {Acta Materialia},
	pages = {72--85},
	publisher = {Elsevier},
	title = {An analysis of the influence of grain size on the strength of FCC polycrystals by means of computational homogenization},
	volume = {148},
	year = {2018}}

@article{molkeri2022importance,
	author = {Molkeri, Abhilash and Khatamsaz, Danial and Couperthwaite, Richard and James, Jaylen and Arr{\'o}yave, Raymundo and Allaire, Douglas and Srivastava, Ankit},
	journal = {Acta Materialia},
	pages = {117471},
	publisher = {Elsevier},
	title = {On the importance of microstructure information in materials design: PSP vs PP},
	volume = {223},
	year = {2022}}

@article{generale2024inverse,
	author = {Generale, Adam P and Robertson, Andreas E and Kelly, Conlain and Kalidindi, Surya R},
	journal = {Acta Materialia},
	pages = {119877},
	publisher = {Elsevier},
	title = {Inverse stochastic microstructure design},
	volume = {271},
	year = {2024}}

@article{lieberman2016microstructural,
	author = {Lieberman, Evan J and Lebensohn, Ricardo A and Menasche, David B and Bronkhorst, Curt A and Rollett, Anthony D},
	journal = {Acta Materialia},
	pages = {270--280},
	publisher = {Elsevier},
	title = {Microstructural effects on damage evolution in shocked copper polycrystals},
	volume = {116},
	year = {2016}}

@article{vogler2008heterogeneous,
	author = {Vogler, TJ and Clayton, John D},
	journal = {Journal of the Mechanics and Physics of Solids},
	number = {2},
	pages = {297--335},
	publisher = {Elsevier},
	title = {Heterogeneous deformation and spall of an extruded tungsten alloy: plate impact experiments and crystal plasticity modeling},
	volume = {56},
	year = {2008}}

@article{remington2018spall,
	author = {Remington, TP and Hahn, EN and Zhao, S and Flanagan, R and Mertens, JCE and Sabbaghianrad, S and Langdon, Terence G and Wehrenberg, CE and Maddox, BR and Swift, DC and others},
	journal = {Acta Materialia},
	pages = {313--329},
	publisher = {Elsevier},
	title = {Spall strength dependence on grain size and strain rate in tantalum},
	volume = {158},
	year = {2018}}

@article{hufnagel2026aimd,
	author = {Hufnagel, Todd C and Addepalli, Pranav and Bhattacharjee, Anuruddha and Berlia, Rohit and El-Awady, Jaafar and Elbert, David and Graham-Brady, Lori and Krieger, Axel and Neralla, Harichandana and Nkansah-Mahaney, T Joseph and others},
	journal = {arXiv preprint arXiv:2603.06835},
	title = {AIMD-L: An automated laboratory for high-throughput characterization of structural materials for extreme environments},
	year = {2026}}

@article{wanchoo2026high,
	author = {Wanchoo, Piyush and Berlia, Rohit and Weihs, Timothy and Ramesh, KT},
	journal = {Journal of the Mechanics and Physics of Solids},
	pages = {106523},
	publisher = {Elsevier},
	title = {High-throughput dynamic experiments: The statistics of spall failure at ultra-high strain rates},
	year = {2026}}

@article{barmak2026machine,
	author = {Barmak, K and Rickman, JM},
	journal = {Acta Materialia},
	pages = {122041},
	publisher = {Elsevier},
	title = {Machine learning for materials microstructures: A survey of applications and methodologies},
	year = {2026}}

@article{liu2015machine,
	author = {Liu, Ruoqian and Yabansu, Yuksel C and Agrawal, Ankit and Kalidindi, Surya R and Choudhary, Alok N},
	journal = {Integrating Materials and Manufacturing Innovation},
	number = {1},
	pages = {192--208},
	publisher = {Springer},
	title = {Machine learning approaches for elastic localization linkages in high-contrast composite materials},
	volume = {4},
	year = {2015}}

@article{oishi2017computational,
	author = {Oishi, Atsuya and Yagawa, Genki},
	journal = {Computer Methods in Applied Mechanics and Engineering},
	pages = {327--351},
	publisher = {Elsevier},
	title = {Computational mechanics enhanced by deep learning},
	volume = {327},
	year = {2017}}

@article{bock2019review,
  title={A review of the application of machine learning and data mining approaches in continuum materials mechanics},
  author={Bock, Frederic E and Aydin, Roland C and Cyron, Christian J and Huber, Norbert and Kalidindi, Surya R and Klusemann, Benjamin},
  journal={Frontiers in Materials},
  volume={6},
  pages={110},
  year={2019},
  publisher={Frontiers}
}

@article{yang2019establishing,
	author = {Yang, Zijiang and Yabansu, Yuksel C and Jha, Dipendra and Liao, Wei-keng and Choudhary, Alok N and Kalidindi, Surya R and Agrawal, Ankit},
	journal = {Acta Materialia},
	pages = {335--345},
	publisher = {Elsevier},
	title = {Establishing structure-property localization linkages for elastic deformation of three-dimensional high contrast composites using deep learning approaches},
	volume = {166},
	year = {2019}}

@article{bhaduri2022stress,
	author = {Bhaduri, Anindya and Gupta, Ashwini and Graham-Brady, Lori},
	journal = {Composites Part B: Engineering},
	pages = {109879},
	publisher = {Elsevier},
	title = {Stress field prediction in fiber-reinforced composite materials using a deep learning approach},
	volume = {238},
	year = {2022}}

@article{gupta2023accelerated,
	author = {Gupta, Ashwini and Bhaduri, Anindya and Graham-Brady, Lori},
	journal = {Mechanics of Materials},
	pages = {104709},
	publisher = {Elsevier},
	title = {Accelerated multiscale mechanics modeling in a deep learning framework},
	volume = {184},
	year = {2023}}

@article{chen2025predicting,
	author = {Chen, Zeping and Yacouti, Marwa and Shakiba, Maryam and Wang, Jian-Xun and Luo, Tengfei and Varshney, Vikas},
	journal = {Theoretical and Applied Mechanics Letters},
	pages = {100634},
	publisher = {Elsevier},
	title = {Predicting stress and damage in carbon fiber-reinforced composites deformation process using composite U-net surrogate model},
	year = {2025}}

@article{buzzy2026polymicros,
	author = {Buzzy, Michael O and Robertson, Andreas E and Chen, Peng and Kalidindi, Surya R},
	journal = {Acta Materialia},
	pages = {122597},
	publisher = {Elsevier},
	title = {PolyMicros: A foundation model for polycrystalline material structure},
	year = {2026}}

@article{bhaduri2021efficient,
	author = {Bhaduri, Anindya and Gupta, Ashwini and Olivier, Audrey and Graham-Brady, Lori},
	journal = {Computational Materials Science},
	pages = {110709},
	publisher = {Elsevier},
	title = {An efficient optimization based microstructure reconstruction approach with multiple loss functions},
	volume = {199},
	year = {2021}}

@article{gao2025recent,
	author = {Gao, Feiyang and Zhang, Weiwen and Liu, Xiaotao and Luo, Xuan and Wang, Zhi and Li, Ning and Liu, Lehua},
	journal = {Journal of Materials Science},
	number = {37},
	pages = {16655--16683},
	publisher = {Springer},
	title = {Recent progress in titanium alloys development via machine learning and high-throughput preparation},
	volume = {60},
	year = {2025}}

@article{zuo2026strength,
	author = {Zuo, MY and Zeng, YH and Sun, K and Sun, J and Wang, M},
	journal = {Materials \& Design},
	pages = {115775},
	publisher = {Elsevier},
	title = {Strength prediction and microstructure quantification of Ti-6Al-4V alloy by selective laser melting based on machine learning},
	year = {2026}}

@article{qin2026multidimensional,
	author = {Qin, Xu and Wang, Qinghang and Yan, Jianchuan and Yang, Biaobiao and Zhang, Yuhui and Jiang, Bin and LLorca, Javier},
	journal = {Acta Materialia},
	pages = {122159},
	publisher = {Elsevier},
	title = {A multidimensional continuous conditional generative adversarial network to link composition-process-microstructure-properties in Mg alloys},
	year = {2026}}

@article{kovachki2022multiscale,
	author = {Kovachki, Nikola and Liu, Burigede and Sun, Xingsheng and Zhou, Hao and Bhattacharya, Kaushik and Ortiz, Michael and Stuart, Andrew},
	journal = {Mechanics of Materials},
	pages = {104156},
	publisher = {Elsevier},
	title = {Multiscale modeling of materials: Computing, data science, uncertainty and goal-oriented optimization},
	volume = {165},
	year = {2022}}

@article{attari2023towards,
	author = {Attari, Vahid and Khatamsaz, Danial and Allaire, Douglas and Arroyave, Raymundo},
	journal = {Acta Materialia},
	pages = {119204},
	publisher = {Elsevier},
	title = {Towards inverse microstructure-centered materials design using generative phase-field modeling and deep variational autoencoders},
	volume = {259},
	year = {2023}}

@article{kalidindi2020feature,
	author = {Kalidindi, Surya R},
	journal = {Journal of Applied Physics},
	number = {4},
	publisher = {AIP Publishing},
	title = {Feature engineering of material structure for AI-based materials knowledge systems},
	volume = {128},
	year = {2020}}

@article{toda2023surrogate,
	author = {Toda, Hiroyuki and Li, Han and Batres, Rafael and Hirayama, Kyosuke and Fujihara, Hiro},
	journal = {Acta Materialia},
	pages = {119188},
	publisher = {Elsevier},
	title = {Surrogate-based optimization of microstructural features of structural materials},
	volume = {257},
	year = {2023}}

@article{liu2015predictive,
	author = {Liu, Ruoqian and Kumar, Abhishek and Chen, Zhengzhang and Agrawal, Ankit and Sundararaghavan, Veera and Choudhary, Alok},
	journal = {Scientific reports},
	number = {1},
	pages = {11551},
	publisher = {Nature Publishing Group UK London},
	title = {A predictive machine learning approach for microstructure optimization and materials design},
	volume = {5},
	year = {2015}}

@article{hendrycks2016baseline,
	author = {Hendrycks, Dan and Gimpel, Kevin},
	journal = {arXiv preprint arXiv:1610.02136},
	title = {A baseline for detecting misclassified and out-of-distribution examples in neural networks},
	year = {2016}}

@article{rabanser2019failing,
	author = {Rabanser, Stephan and G{\"u}nnemann, Stephan and Lipton, Zachary},
	journal = {Advances in Neural Information Processing Systems},
	title = {Failing loudly: An empirical study of methods for detecting dataset shift},
	volume = {32},
	year = {2019}}

@article{perdikaris2017nonlinear,
	author = {Perdikaris, Paris and Raissi, Maziar and Damianou, Andreas and Lawrence, Neil D and Karniadakis, George Em},
	journal = {Proceedings of the Royal Society A: Mathematical, Physical and Engineering Sciences},
	number = {2198},
	publisher = {The Royal Society},
	title = {Nonlinear information fusion algorithms for data-efficient multi-fidelity modelling},
	volume = {473},
	year = {2017}}

@article{gretton2012kernel,
	author = {Gretton, Arthur and Borgwardt, Karsten M and Rasch, Malte J and Sch{\"o}lkopf, Bernhard and Smola, Alexander},
	journal = {The journal of machine learning research},
	number = {1},
	pages = {723--773},
	publisher = {JMLR. org},
	title = {A kernel two-sample test},
	volume = {13},
	year = {2012}}

@article{quey2011large,
	author = {Quey, Romain and Dawson, Paul R and Barbe, Fabrice},
	journal = {Computer Methods in Applied Mechanics and Engineering},
	number = {17-20},
	pages = {1729--1745},
	publisher = {Elsevier},
	title = {Large-scale 3D random polycrystals for the finite element method: Generation, meshing and remeshing},
	volume = {200},
	year = {2011}}

@article{saha2026numerical,
	author = {Saha, Indrashish and Graham-Brady, Lori},
	journal = {Computer Methods in Applied Mechanics and Engineering},
	pages = {118493},
	publisher = {Elsevier},
	title = {Numerical and data-driven modeling of spall failure in polycrystalline ductile materials},
	volume = {448},
	year = {2026}}

@book{antoun2006spall,
  title={Spall fracture},
  author={Antoun, Tarabay and Seaman, Lynn and Curran, Donald R and Kanel, Gennady I and Razorenov, Sergey V and Utkin, Alexander V},
  year={2006},
  publisher={Springer Science \& Business Media}
}

@article{saha2024prediction,
	author = {Saha, Indrashish and Gupta, Ashwini and Graham-Brady, Lori},
	journal = {Computer Methods in Applied Mechanics and Engineering},
	pages = {116816},
	publisher = {Elsevier},
	title = {Prediction of local elasto-plastic stress and strain fields in a two-phase composite microstructure using a deep convolutional neural network},
	volume = {421},
	year = {2024}}

@article{di2023neural,
	author = {Di Leoni, Patricio Clark and Lu, Lu and Meneveau, Charles and Karniadakis, George Em and Zaki, Tamer A},
	journal = {Journal of Computational Physics},
	pages = {111793},
	publisher = {Elsevier},
	title = {Neural operator prediction of linear instability waves in high-speed boundary layers},
	volume = {474},
	year = {2023}}

@article{pyzer2022accelerating,
	author = {Pyzer-Knapp, Edward O and Pitera, Jed W and Staar, Peter WJ and Takeda, Seiji and Laino, Teodoro and Sanders, Daniel P and Sexton, James and Smith, John R and Curioni, Alessandro},
	journal = {npj Computational Materials},
	number = {1},
	pages = {84},
	publisher = {Nature Publishing Group UK London},
	title = {Accelerating materials discovery using artificial intelligence, high performance computing and robotics},
	volume = {8},
	year = {2022}}

\end{document}